\documentclass[longauth]{aa} 
\usepackage{graphicx}
\usepackage{txfonts}
\usepackage{rotating}
\usepackage{xcolor}
\usepackage[hidelinks]{hyperref}
\usepackage{mathtools}

\pdfoutput=1

\newcommand{\mj}{miniJPAS}
\newcommand{\link}{\href{https://www.j-pas.org/ancillarydata/minijpas\_amico\_galaxy\_clusters}{www.j-pas.org/ancillarydata/minijpas\_amico\_galaxy\_clusters}}

\begin{document} 

\title{The miniJPAS survey:}
\subtitle{clusters and galaxy groups detection with AMICO}
\authorrunning{M. Maturi and J-PAS collaboration}
\titlerunning{AMICO-miniJPAS cluster catalogue}


\author{
    M. Maturi\inst{1,2}
    \and
    A. Finoguenov\inst{3}
    \and
    P. A. A. Lopes\inst{\ref{OV}}
    \and
    R.~M.~Gonz\'alez Delgado\inst{\ref{iaa}} 
    \and
    R. A. Dupke\inst{\ref{ON},\ref{DAUM},\ref{DPA}}
    \and
    E. S. Cypriano\inst{\ref{astrosaopaulo}}
    \and
    E. R. Carrasco\inst{\ref{GEM}}
    \and
    J.M. Diego\inst{\ref{IFCA}}
    \and
    M. Penna-Lima\inst{\ref{UnB}} 
    \and
    J. M. V\'ilchez\inst{\ref{iaa}}
    \and
    L. Moscardini\inst{\ref{Bologna1},\ref{Bologna2},\ref{Bologna3}}
    \and
    V. Marra\inst{\ref{NucleoES},\ref{INAF_TR},\ref{IFPU}}
    \and
    S. Bonoli\inst{\ref{teruel},\ref{DIPC},\ref{Ikerb}}
    \and
    J.E. Rodr\'iguez-Mart\'in\inst{\ref{iaa}}
    \and
    A. Zitrin\inst{\ref{BenGu}}
    \and
    I. M\'arquez\inst{\ref{iaa}}
    \and
    A.~Hern\'an-Caballero\inst{\ref{teruel}}
    \and
    Y. Jim\'enez-Teja\inst{\ref{iaa}}    
    \and
    R.~Abramo\inst{\ref{fisicasaopaulo}}
    \and
    J.~Alcaniz\inst{\ref{ON}}
    \and
    N.~Benitez\inst{\ref{iaa}}
    \and
    S.~Carneiro\inst{\ref{fisicabahia}}
    \and
    J.~Cenarro\inst{\ref{teruel}}
    \and
    D.~Cristóbal-Hornillos\inst{\ref{teruel}}
    \and
    A.~Ederoclite\inst{\ref{teruel}}
    \and
    C.~López-Sanjuan\inst{\ref{teruel}}
    \and
    A.~Marín-Franch\inst{\ref{teruel}}
    \and
    C.~Mendes~de~Oliveira\inst{\ref{astrosaopaulo}}
    \and
    M.~Moles\inst{\ref{teruel}}
    \and
    L.~Sodré~Jr.\inst{\ref{astrosaopaulo}}
    \and
    K.~Taylor\inst{\ref{instruments4}}
    \and
    J.~Varela\inst{\ref{teruel}}
    \and
    H. V\'azquez Rami\'o\inst{\ref{teruel}}
    \and
    J.A. Fern\'andez-Ontiveros\inst{\ref{teruel}}
}

\institute{
    Zentrum f\"ur Astronomie, Universitat\"at Heidelberg, Philosophenweg 12, D-69120 Heidelberg, Germany\\
    \email{maturi@uni-heidelberg.de}
\and
    Institute for Theoretical Physics, Philosophenweg 16, D-69120 Heidelberg, Germany
\and
    Department of Physics, University of Helsinki, P.O. Box 64, FI-00014 Helsinki, Finland
\and
    Observat\'orio do Valongo, Universidade Federal do Rio de Janeiro, Ladeira do Pedro Ant\^onio 43, Rio de Janeiro, RJ, 20080-090, Brazil\label{OV}
\and
    Instituto de Astrof\'isica de Andaluc\'ia (IAA-CSIC), Glorieta de la Astronomía s/n, E-18008 Granada, Spain\label{iaa}
\and
    Observatório Nacional, Rua General José Cristino, 77, São Cristóvão, 20921-400, Rio de Janeiro, RJ, Brazil\label{ON}
\and
    Department of Astronomy, University of Michigan, 311 West Hall, 1085 South University Ave., Ann Arbor, USA\label{DAUM}
\and
    Department of Physics and Astronomy, University of Alabama, Box 870324, Tuscaloosa, AL, USA\label{DPA}
\and
    Departamento de Astronomia, Instituto de Astronomia, Geof\'isica e Ci\^encias Atmosf\'ericas da USP, Cidade Universit\'aria, \\ 05508-900, S\~ao Paulo, SP, Brazil\label{astrosaopaulo}
\and
    Gemini Observatory,NSF's NOIRLab, Casilla 603, La Serena 1700000, Chile\label{GEM}
\and
    Instituto de F\'isica de Cantabria (CSIC-UC). Avda. Los Castros s/n. 39005 Santander, Spain\label{IFCA}
\and
    Universidade de Bras\'ilia, Instituto de F\'isica, Caixa Postal 04455, Bras\'ilia, DF, 70919-970, Brazil\label{UnB}
\and
    Dipartimento di Fisica e Astronomia “Augusto Righi” - Alma Mater Studiorum Universit\`a di Bologna, via Piero Gobetti 93/2, I-40129 Bologna, Italy\label{Bologna1}
\and
    INAF-Osservatorio di Astrofisica e Scienza dello Spazio di Bologna, Via Piero Gobetti 93/3, I-40129 Bologna, Italy\label{Bologna2}
\and
    INFN-Sezione di Bologna, Viale Berti Pichat 6/2, I-40127 Bologna, Italy\label{Bologna3}
\and
    Núcleo de Astrofísica e Cosmologia \& Departamento de Física, Universidade Federal do Espírito Santo, 29075-910, Vitória, ES, Brazil\label{NucleoES}
\and
    INAF -- Osservatorio Astronomico di Trieste, via Tiepolo 11, 34131 Trieste, Italy\label{INAF_TR}
\and
    IFPU -- Institute for Fundamental Physics of the Universe, via Beirut 2, 34151, Trieste, Italy\label{IFPU}
\and
    Centro de Estudios de F\'isica del Cosmos de Arag\'on (CEFCA), Unidad Asociada al CSIC, Plaza San Juan, 1, E-44001 Teruel, Spain\label{teruel}
\and
    Donostia International Physics Center (DIPC), Manuel Lardizabal Ibilbidea, 4, San Sebasti\'an, Spain\label{DIPC}
\and
    Ikerbasque, Basque Foundation for Science, 48013 Bilbao, Spain\label{Ikerb}
\and
    Physics Department, Ben-Gurion University of the Negev, P.O. Box 653, Be'er-Sheva 84105, Israel\label{BenGu}
\and
    Departamento de F\'isica Matem\'atica, Instituto de F\'{\i}sica, Universidade de S\~ao Paulo, Rua do Mat\~ao, 1371, CEP 05508-090, S\~ao Paulo, Brazil \label{fisicasaopaulo}
\and
    Instituto de F\'isica, Universidade Federal da Bahia, 40210-340, Salvador, BA, Brazil\label{fisicabahia}
\and
    Instruments4, 4121 Pembury Place, La Canada Flintridge, CA 91011, U.S.A\label{instruments4}
}

\date{\today}

\abstract
   {Samples of galaxy clusters allow us to better understand the physics at play in galaxy formation and to constrain cosmological models once their mass, position (for clustering studies) and redshift are known. In this context, large optical data sets play a crucial role.
   }
   {We investigate the capabilities of the Javalambre-Physics of the Accelerating Universe Astrophysical Survey (J-PAS) in detecting and characterizing galaxy groups and clusters. We analyze the data of the \mj survey, obtained with the JPAS-{\it Pathfinder} camera and covering 1~deg$^2$ centered on the AEGIS field to the same depths and with the same 54 narrow band plus 2 broader band near-UV and near-IR filters anticipated for the full J-PAS survey.}
   {We use the Adaptive Matched Identifier of Clustered Objects (AMICO) to detect and characterize groups and clusters of galaxies down to $S/N=2.5$ in the redshift range $0.05<z<0.8$.}
   {We detect $80$, $30$ and $11$ systems with signal-to-noise ratio larger than $2.5$, $3.0$ and $3.5$, respectively, down to $\sim 10^{13}\,M_\odot/h$. We derive mass-proxy scaling relations based on Chandra and XMM-Newton X-ray data for the signal amplitude returned by AMICO, the intrinsic richness and a new proxy that incorporates the galaxies' stellar masses. The latter proxy is made possible thanks to the J-PAS filters and shows a smaller scatter with respect to the richness. We fully characterize the sample and use AMICO to derive a probabilistic membership association of galaxies to the detected groups that we test against spectroscopy. We further show how the narrow band filters of J-PAS provide a gain of up to $100$\% in signal-to-noise ratio in detection and an uncertainty on the redshift of clusters of only $\sigma_z=0.0037(1+z)$ placing J-PAS in between broadband photometric and spectroscopic surveys.}
   {
   The performances of AMICO and J-PAS with respect to mass sensitivity, mass-proxies quality and redshift accuracy will allow us to derive cosmological constraints not only based on cluster counts, but also clustering of galaxy clusters.}

\keywords{galaxies: clusters: general / galaxies: evolution / galaxies: luminosity function, mass function}

\maketitle

\section{Introduction}

The formation of structures in the universe is extremely sensitive to cosmic expansion captured by the cosmological parameters such as for example the matter density parameter $\Omega_m$, the power spectrum normalisation $\sigma_8$ and the equation of state of dark energy whenever high-redshift data sets are available. Notable examples are the study of the large scale structure through galaxy clustering \citep{kaiser87,caccaito13,vakili20,padey21}, cosmic shear \citep{Blandford1991,Bernardeau2010,hildebrandt2017}, baryonic acoustic oscillations \citep{Hu1996,blake11,hinton17,costa19,duMasDesBourdoux20} and cosmic microwave background measurements \citep{pettorino12,hinshaw13,arnaud16}, which probe the linear regime of the growth of structures. On the other end, galaxy clusters sit in the exponential tails of the cosmic mass function being the largest gravitationally bound structures and as such are extremely sensitive to the background cosmology, making them ideal probes for cosmological analysis \citep{allen11,benson13,planckcoll15,bocquet19,abbott20,finoguenov20,iderchithmam20,costanzi21,marulli21,giocoli21,ingoglia22,lesci22,lesci22b,garrel22,chiu22}.

\begin{figure*}[h!]
    \includegraphics[width=0.48\textwidth]{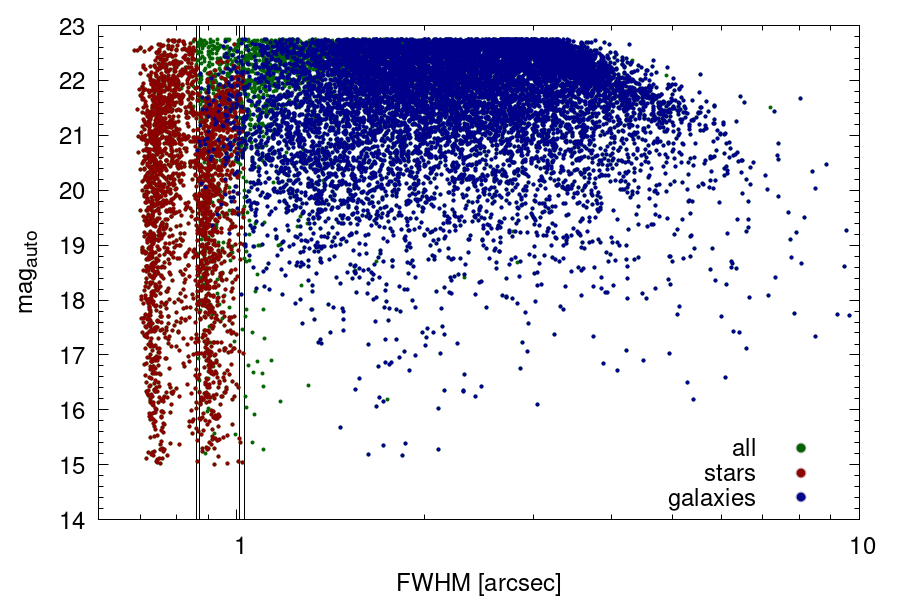}
    \includegraphics[width=0.48\textwidth]{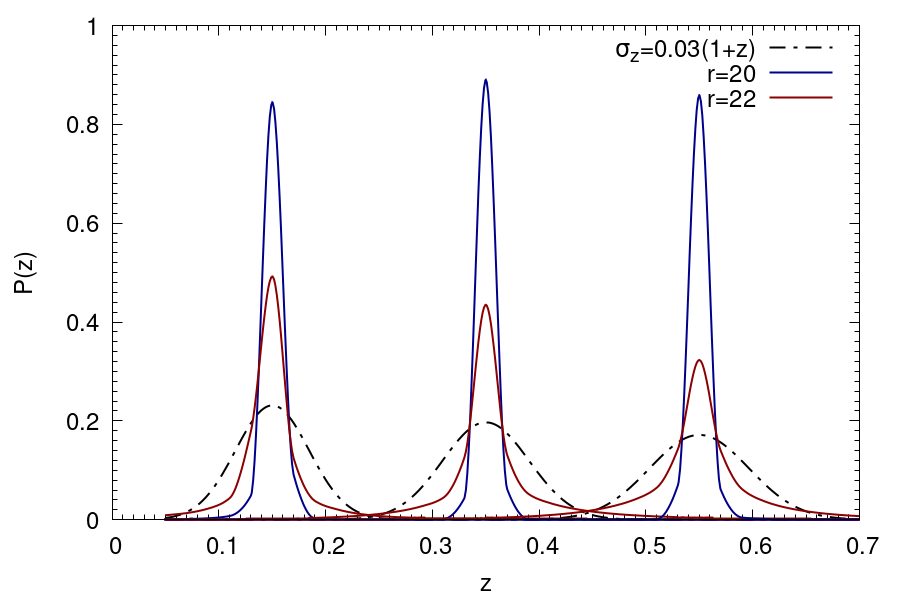}
    \caption{Left panel: $r$-band magnitude v.s. FWHM of the galaxies detected in \mj and used for the analysis (in blue). The objects with a Bayesian probability of being stars/galaxies larger/smaller than 50\% are shown in red/blue. All objects, in green, with a FWHM smaller than the PSF size of each individual pointing plus a tolerance of 0.11 arcsec (indicated by the vertical lines) have been rejected regardless their classification. Right panel: average redshift probability distribution of galaxies as measured by AMICO with a peak $P(z)$ located at three different redshifts and for two different magnitudes: $r=20$ and $r=22$ in blue and red, respectively. This estimated is produced by AMICO through the input photo-$z$s and is used in the cluster model construction. The black dashed lines show the typical uncertainty of photo-$z$s based on broad band photometry.}
    \label{fig:galaxy_stat}
\end{figure*}

For this reason, large campaigns aiming at the detection of large samples of galaxy clusters have been conducted in the last decades in the millimetric wavelengths \citep{bleem15,zubeldia19,hilton20}, in the optical regime \citep{Rykoff14,rykoff16,maturi19} and in the X-rays \citep{bohringer04,vikhlinin09,finoguenov20}. Further data sets are going to be provided in the near future by wide field surveys such as eROSITA \citep{merloni12,pillepich12,kafer20,ota22}, LSST \citep{LSST2009}, Euclid \citep{Sartoris16,scaramella22}, 4MOST \citep{dejong14} and J-PAS \citep{benitez14}. In particular, optical surveys are gaining momentum thanks to the availability of the complete wavelength range filling set of narrow band filters with high uniformity and transmission \citep{martin-franch12} and refined techniques for the estimation of photometric redshifts \citep{Arnouts11,Bilicki18,benitez20,hernanCaballero21}, allowing one to detect a large number of clusters with masses smaller compared to those detected with Sunyaev Zel'dovich observations, and redshifts higher when compared to those from typical of X-ray observations. In addition, they directly provide  accurate redshift and membership estimates without the need of a subsequent follow-up.

Several algorithms for cluster detection have been proposed and used in the recent years \citep[see e.g.][]{Farrens11,licitera2016,rykoff16,adam19}, exploiting both different cluster optical properties and different methodologies. In this work we discuss the application of the Adaptive Matched Identifier of Clustered Objects \citep[AMICO,][]{bellagamba18b,maturi19} to the \mj data set \citep{bonoli20} to investigate the prospects for the detection of galaxy clusters in the Javalambre Physics of the Accelerating Universe Astrophysical Survey \citep[J-PAS,][]{benitez14}. We construct the cluster sample and we determine mass-proxy scaling relations based on the X-ray mass estimates using Chandra and XMM-Newton data. Furthermore, we provide a catalogue of cluster members based on the probabilistic membership association provided by AMICO itself, we identify the brightest group galaxy (BGG), and we illustrate the power of the 56 J-PAS filters \citep{martin-franch12}. To construct the cluster model used for the cluster detection, we assumed the following cosmological parameters $\Omega_m=0.3$, $\Omega_\Lambda=0.7$ and $h=0.7$.

The plan of the paper is as follows. In Section~\ref{sec:data} we summarize the properties of the data set and in Section~\ref{sec:cluster-sample} we describe the detection algorithm and characterize the cluster sample through X-ray and spectroscopic data. In Section~\ref{sec:members-bcg} we present the samples of cluster members and Bright Group Galaxies we produced in the analysis. In Section~\ref{sec:spectral} we characterize the cluster sample through spectroscopic redshifts and investigate the accuracy of the probabilistic memberships provided by AMICO. The X-ray counterparts of our sample are discussed in Section~\ref{sec:xrays},  while in Section~\ref{sec:impact-narrow-filter} we discuss the gain provided by the narrow band J-PAS filters with respect to broad band ones. Finally, we give our conclusions in Section~\ref{sec:conclusions}.

\section{The data set}\label{sec:data}

In our study we consider data from the \mj survey \citep{bonoli20}, a $\sim 1$~deg$^2$ survey covering the AEGIS field along the Extended Groth Strip and serving as a test of the upcoming much wider J-PAS survey. The data have been obtained with the $2.55$m telescope JST at the Javalambre Astrophysical Observatory and an interim camera (J-PAS {\it  Pathfinder} camera) equipped with a $9k\times9k$ CCD covering a 0.3~deg$^2$ field-of-view with a pixel size of $0.23^{\prime\prime}$. The distinctive features of this survey are its 56 J-PAS filters which, on top of the $u,g,r,i$ SDSS broad band filters, comprise 54 narrow band (FWHM$\sim145\AA$) and two broader filters extending to the near-UV and the near-infrared.

\begin{figure*}[h!]
    \includegraphics[width=0.99\textwidth]{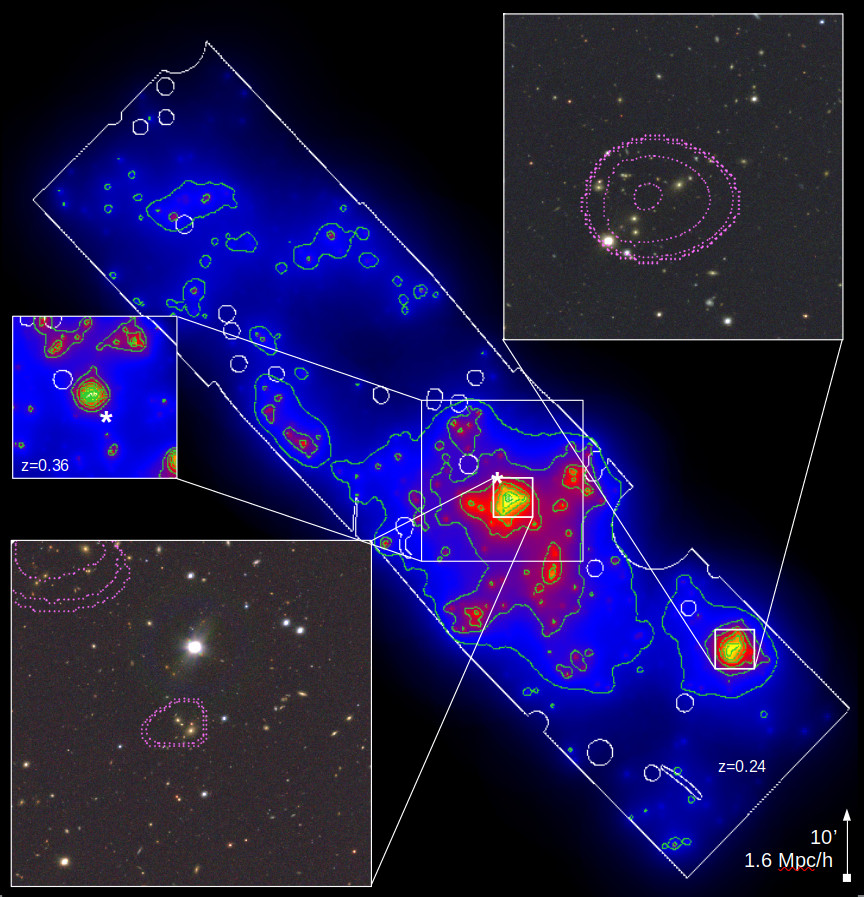}
    \caption{In the background, illustration of the AMICO's response amplitude for a redshift slice at $z=0.24$. The overdensities highlighted by AMICO are clearly visible as strong enhancements over the background. The survey footprint and masked areas are indicated in white. The two boxes with the \mj images show the two most significant detections at that redshift together with X-ray emission isocontours. The top left box shows the response of AMICO around the main central structure at $z=0.24$, but for redshift $z=0.36$, where another cluster is clearly visible. Note the lack of cross-contamination between redshift slices. Also the cluster on the right hand side is a case of chance alignment between two structures, see left panel of Fig.~\ref{fig:amico-detections_picture}. The asterisks in the figures indicate the position of the cluster visible in the other redshif slice and displaced nearly along the same line of side.}
    \label{fig:amico-amplitude-map}
\end{figure*}

For constructing the cluster sample, we selected all objects in the magnitude range $15.0<r<22.75$, with a SExtractor flag smaller or equal to 3, a star classification probability below $P_{star} < 0.5$ and a  $FWHM$ in the r-band images larger than $0.86$, $0.87$, $1.01$ or $1.03$~arcsec according to the criteria $FWHM_{galaxy}>FWHM_{PSF}+0.11$ to avoid any possible contamination due to misclassified stars. The $0.11$ additive factor was set euristically to avoid the locus in which star are located. Here, $FWHM_{PSF}$ is the $FWHM$ of the Point Spread Function (PSF) of each individual pointing. We adopted a relatively relaxed constraint on $P_{star}$ because we used it in combination with the cut off at the minimum $FWHM$ values. We also tested a more conservative value of $P_{star} < 0.95$ both with and without $FWHM$ cut off, and the final results do not change significantly. In Fig.~\ref{fig:galaxy_stat} we show the main properties of the photometric sample comprising stars and galaxies. The left panel of the figure shows the distribution of objects classified as stars (in red), galaxies (in blue) and all objects (in green) with a FWHM smaller than the PSF size of each individual pointing plus a tolerance of 0.11 arcsec (indicated by the vertical lines) that have been rejected regardless their classification (in green). The presence of the two regions mostly populated by objects classified as stars at low FWHM shows that the PSF quality for the different pointings is bimodal. This data set provides an ideal playground to test the survey strategy for the detection of galaxy clusters in the final J-PAS survey that will cover a $8000$ times larger area.

\begin{figure*}[h!]
    \includegraphics[width=0.33\textwidth]{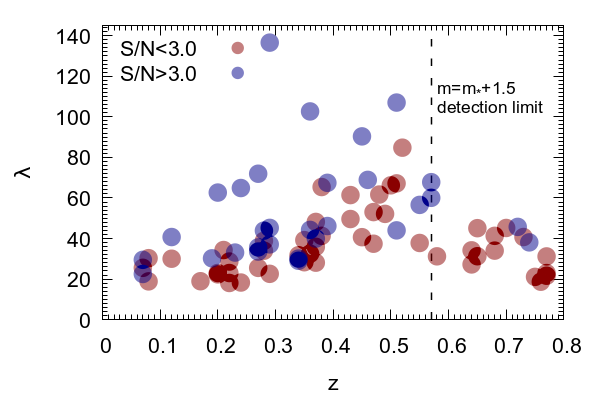}
    \includegraphics[width=0.33\textwidth]{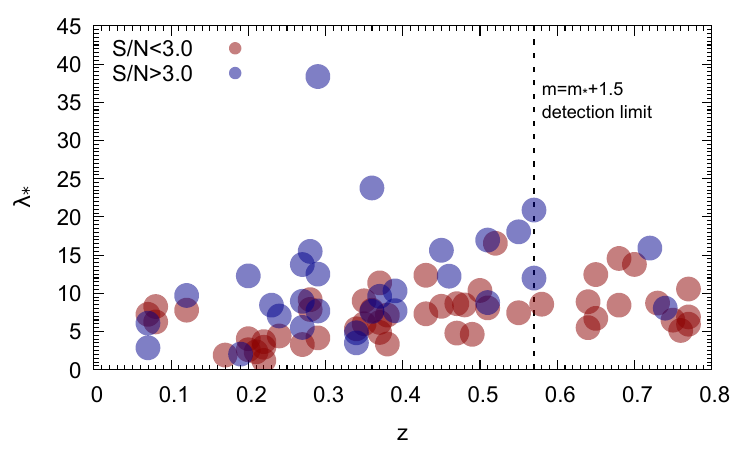}
    \includegraphics[width=0.33\textwidth]{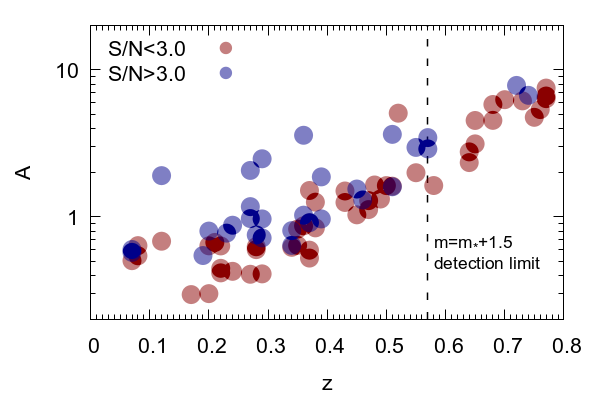}
    \caption{The observed richness $\lambda$ (left panel), and the intrinsic richness $\lambda_*$ (central panel) as a function of redshift obtained with AMICO. For redshifts larger than $z=0.56$, the cut off magnitude $m_{cut}=m_*+1.5$, on which $\lambda_*$ is based, falls below the detection limit and the richness is underestimated as it is clear from the lack of rise in the $\lambda_*$ of detections with $z>0.56$. The right panel shows the amplitude, $A$, of the detections. Blue and red symbols refer to detections with $S/N>3$ and $S/N<3$, respectively.}
    \label{fig:amico-detections_proxies}
\end{figure*}

\begin{figure*}[h!]
    \includegraphics[width=0.33\textwidth]{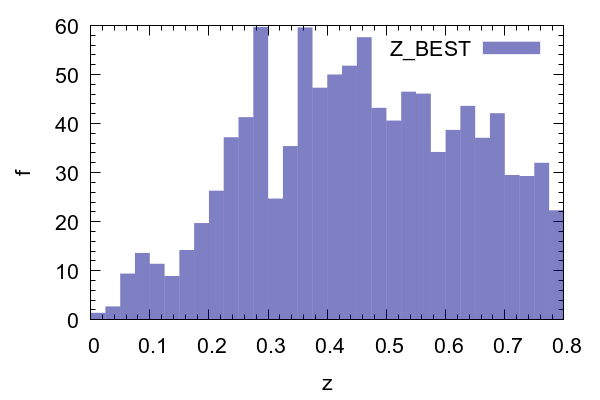}
    \includegraphics[width=0.33\textwidth]{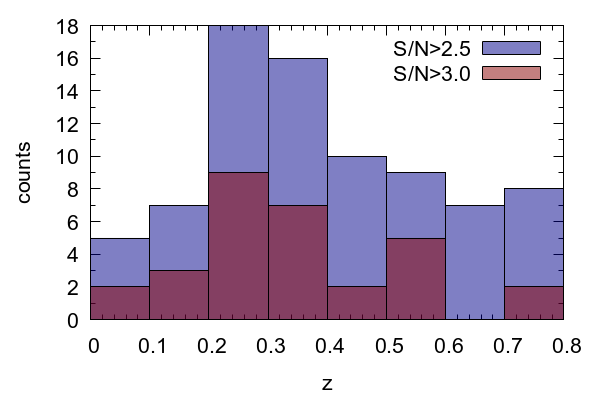}
    \includegraphics[width=0.33\textwidth]{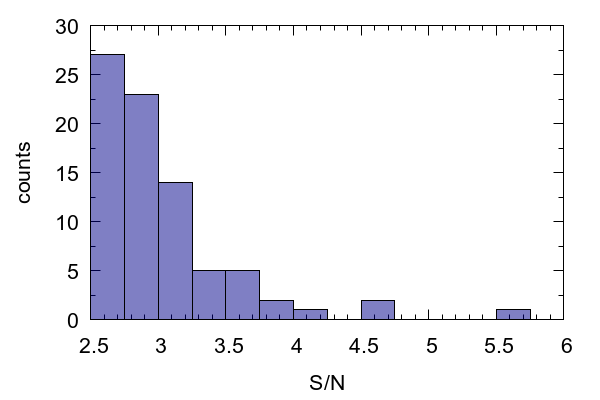}
    \caption{The redshift distribution of the miniJPAS galaxies (left panel) and of the AMICO cluster detections (central panel; blue and red histrograms for detections with $S/N>2.5$ and $S/N>3$, respectively). The large fluctuations in the number of galaxies and clusters are physical and are due to large scale structure of the field as confirmed by the spectroscopic redshifts \citep{bonoli20}. The right panel shows the distribution of signal-to-noise ratio of the detections.}
    \label{fig:amico-detections_z_distri}
\end{figure*}

Thanks to the availability of the 54 narrow band and 2 broader band near-UV and near-IR filters, the photo-$z$s have an accuracy with no comparison with respect to other photometric surveys and place \mj in between low resolution spectroscopy and typical broad band photometry. A detailed study of the photo-$z$ characteristics is presented in \cite{hernanCaballero21}. In the right panel of Fig.~\ref{fig:galaxy_stat} we show the average redshift probability distribution of galaxies of magnitude $r=20$ and $r=22$ located at different redshifts. The average $P(z)$ is produced by AMICO through the input photo-$z$s and is used to define the cluster model as discussed in \cite{maturi19}. As an illustration, we show the typical photo-$z$ uncertainty of data sets based on broad band filters, here we assumed $\sigma_z = 0.03 (1+z)$ (dashed line). The high accuracy of J-PAS photo-$z$s provides higher sensitivity towards clusters of smaller mass aided by the increase in compactness of cluster members along the redshift direction with decreasing mass.
The improved accuracy on photo-$z$s results also in a better estimation of the clusters richness, which in turn improves the cosmology-relevant mass estimation.  


\section{The galaxy cluster sample}\label{sec:cluster-sample}

In this section we summarize the main concepts behind the detection algorithm used in this analysis. We describe the properties of the cluster candidates obtained with the photo-$z$s based on the 56 J-PAS filters.

\subsection{Detection of galaxy groups and clusters with AMICO}\label{sec:amico}

\begin{table*}
    \centering
    \begin{tabular*}{\textwidth}{l@{\extracolsep{\fill}}llr}
        \hline
        \multicolumn{1}{l}{Column} & \multicolumn{1}{l}{Unit} & \multicolumn{1}{c}{Description} & \multicolumn{1}{r}{Example} \\
        \hline
 \texttt{SURVEY} (1) & & survey name & mJP \\
 \texttt{NAME} (2) & & object name & J141426.57+515608 \\
 \texttt{ID} (3) & & unique identification number of the cluster & 7 \\
 \texttt{IDAMICO} (4) & & AMICO identification number & 1001 \\
 \texttt{RA} (5) & deg & position in the sky: x direction (R.A) & 213.6107 \\
 \texttt{DEC} (6) & deg & position in the sky: y direction (Dec) & 51.9358 \\
 \texttt{Z} (7) & redshift & AMICO redshift & 0.290 \\
 \texttt{Z\_ERR} (8) & redhisft & $1\sigma$ uncertainty of AMICO redshift & 0.006 \\
 \texttt{Z\_PROB} (9) & redshift & redshift based on AMICO memberships & 0.290 \\
 \texttt{Z\_PROB\_ERR} (10) & redshift & $1\sigma$ uncertainty based on AMICO memberships & 0.005 \\
 \texttt{Z\_SPEC} (11) & redshift & spectroscopic redshift (when available) & -99 \\
 \texttt{N\_SPEC} (12) & & number of members with spectroscopic redshift & 0 \\
 \texttt{SN} (13)  &  & signal-to-noise ratio of AMP; noise = background + cluster & 5.66 \\
 \texttt{MSKFRC} (14) & & fraction of the cluster inaccessible because of masks & 0.33 \\
 \texttt{R\_MEDIAN} (15) & deg & median separation of galaxies with $P>50$\% & 0.0628 \\ 
 \texttt{R\_50} (16) & deg & radius at which $\int_0^r P(r) \,/ \int_0^{\infty} P(r) = 0.5$ & 0.0838 \\
 \texttt{RICH} (17) & &  apparent richness based on the filter formalism & 170 \\
 \texttt{AMP} (18) & & signal amplitude & 2.48 \\
 \texttt{AMP\_ERR} (19) & & $1\sigma$ error on \texttt{AMP} & 0.44 \\ 
 \texttt{LAMBDA} (20) & & richness based on memberships & 136 \\
 \texttt{LAMBDA\_ERR} (21) & & $1\sigma$ error on \texttt{LAMBDA\_ERR} & 12 \\
 \texttt{LAMBDA\_STAR} (22) & & richness based on memberships and $m_r<m_*+1.5$ and $r<r_{vir}$ & 38.4 \\
 \texttt{LAMBDA\_STAR\_ERR} (23) & & $1\sigma$ error on \texttt{LAMBDA\_STAR} & 6.2 \\
 \texttt{LAMBDA\_MSTAR} (24) & $10^{12} M_\odot$ & total stellar mass based on memberships and $\log_{10}(M_*)>9.5$ & 3.4 \\
 \texttt{LAMBDA\_MSTAR\_ERR} (25) & $10^{12} M_\odot$ & $1\sigma$ error on LAMBDA\_MSTAR & 0.5\\
 \texttt{MASS\_AX} (26) & $10^{13} M_\odot$ & mass based on the amplitude and the X-ray scaling relation  & 0.23 \\
 \texttt{MASS\_AX\_ERR} (27) & $10^{13} M_\odot$ & error of \texttt{MASS\_AX} & 0.02 \\
 \texttt{MASS\_X} (28) & $10^{13} M_\odot$ & mass based on X-rays & 0.87 \\
 \texttt{MASS\_X\_ERR} (29) & $10^{13} M_\odot$ & $1\sigma$ error on \texttt{MASS\_X} & 0.07 \\ 
 \texttt{L\_X} (30) & $10^{42}$erg/s & X-ray luminosity & 14.0 \\
 \texttt{L\_X\_ERR} (31) & $10^{42}$erg/s & $1\sigma$ error on \texttt{L\_X} & 1.8 \\
 \texttt{SN\_X} (32) & & signal-to-noise ratio of the X-ray flux & 7.8 \\
 \texttt{ID\_BGG1} (33) & & identification number of the BGG galaxy, first choice & 1771 \\
 \texttt{ID\_BGG2} (34) & & identification number of the BGG galaxy, second choice & 1639 \\
        \hline
    \end{tabular*}
    \caption{Descriptions of the columns for the AMICO galaxy cluster catalogue. The full VAC is available online at: \link}
    \label{tab:catalogue_columns}
\end{table*}

We use the Adaptive Matched Identifier of Clustered Objects \citep[AMICO,][]{2005A&A...442..851M, bellagamba18} for the detection of galaxy clusters. Thanks to its performances, AMICO has been selected as one of the two algorithms tested and implemented in the official scientific data analysis pipeline of the Euclid space mission of the European Space Agency \citep{adam19} and has been successfully tested and applied to the data of the Kilo Degrees Survey, allowing both cosmological and astrophysical studies \citep{maturi19, bellagamba18b, radovich20, puddu21, tortora20, sereno20, giocoli21, lesci22, smit22, ingoglia22, lesci22}. Its core stands on an optimal linear matched filtering approach in which the data are convolved with a kernel (the so called filter) derived through a constrained minimization approach which minimizes the noise variance under the condition that the estimated signal is unbiased. The two key ingredients of this process are a statistical description of the background noise, $N(\vec{x})$, and a template, $C(\vec{x})$, characterizing the signal of clusters, $S({\vec{x})}=A\,C(\vec{x})$, as a function of the properties $\vec{x}$ and a factor $A$ that scales with the cluster mass, the so called amplitude. The statistical properties of the noise $N(\vec{x})$ can be derived directly from the data, while the nature of the function $C(\vec{x})$ and of the variables $\vec{x}$ depend on the specific application of the algorithm. Here, we restrict ourselves to the case in which $\vec{x}$ includes positions and the r-magnitudes of galaxies only. All details about the model are given in Section~\ref{sec:catalogue} where we describe the full filter formalism.

The result of the filtering applied by AMICO to the data is an estimate of the amplitude,
\begin{equation}
  \label{eq:amplitude}
  A(\vec{\theta}_c,z_c) = \alpha^{-1}(z_c) \sum_{i=1}^{N_{gal}}\frac{C(z_c;\vec{\theta}_i-\vec{\theta}_c,m_i)p_i(z_c)}{N(m_i,z_c)} - B(z_c) \;,
\end{equation}
where $z_c$ and $\vec{\theta}_c$ are the putative redshift and angular position of a cluster, respectively, while $\vec{\theta}_i$,  $m_i$ and $p_i(z)$ are the angular position, the $r$-band magnitude and the photometric redshift distribution of the $i$-th galaxy taken from the input catalogue, respectively. The filter normalization is set by the factor $\alpha$ and the average contribution of the field galaxies to the total signal amplitude is given by $B$. The expected r.m.s. of $A$ is given by
\begin{equation}
  \label{eq:amplitudesigma}
  \sigma_A(\vec{\theta}_c,z_c) = \alpha(z_c)^{-1} + A(\vec{\theta}_c,z_c)\frac{\gamma(z_c)}{\alpha(z_c)^{2}}\;,
\end{equation}
where the first term captures the stochastic fluctuations of the background and the second one the Poissonian fluctuations due to the cluster members.
\begin{equation}
\begin{split}
	B(z_c) &= \alpha^{-1}(z_c) \int C(z_c;\vec \theta - \vec \theta_c, m) q(z_c,z)\ d^2\theta\ dm\ dz  \\
	\alpha(z_c) &= \int \frac {C^2(z_c;\vec \theta - \vec \theta_c, m)\ q_1(m,z_p,z_c) q_2(m,z_c,z_p)} {N(m,z_c)}\ d^2\theta\ dm\ dz_p \\
	\gamma(z_c) &= \int \frac {C^3(z_c;\vec \theta - \vec \theta_c, m)\ q^2_1(m,z_p,z_c) q_2(m,z_c,z_p)} {N^2(m,z_c)}\ d^2\theta\ dm\ dz_p.
\end{split}
\end{equation}
are filter constants expressing the background level, the filter normalization and the contribution to the noise given by the cluster members, respectively. Here, 
\begin{equation}
q(z_c,z) = \left(\sum_{i=1}^{N_\text{gal}} p_i(z_c)  \right)^{-1}  \sum_{i=1}^{N_\text{gal}} p_i(z-z_c+z_\text{peak,i})\ p_i(z_c)
\end{equation}
expresses the typical redshift probability distribution for a galaxy located at redshift $z_c$,
\begin{equation}
  q_1(m,z_p,z_c) = \left(\sum_{z_{\text{peak},i}=z_p} p_i(z_p)  \right)^{-1} \sum_{z_{\text{peak},i}=z_p} p_i(z_p)\ p_i(z_c) ,
\end{equation}
describes the typical $p(z)$ that peaks at $z_p$, and
\begin{equation}
  q_2(m,z_c,z_p) = \left(\sum_{i=1}^{N_\text{gal}} p_i(z_c)  \right)^{-1} \sum_{z_{\text{peak},i}=z_p} p_i(z_c)\ p_i(z_p) ,
\end{equation}
describes the probability distribution for the peak, $z_p$, of a galaxy
located at redshift $z_c$. All details about these expressions are given in \cite{bellagamba18} and \cite{maturi19}.

\begin{figure*}[h!]
    \includegraphics[width=0.46\textwidth]{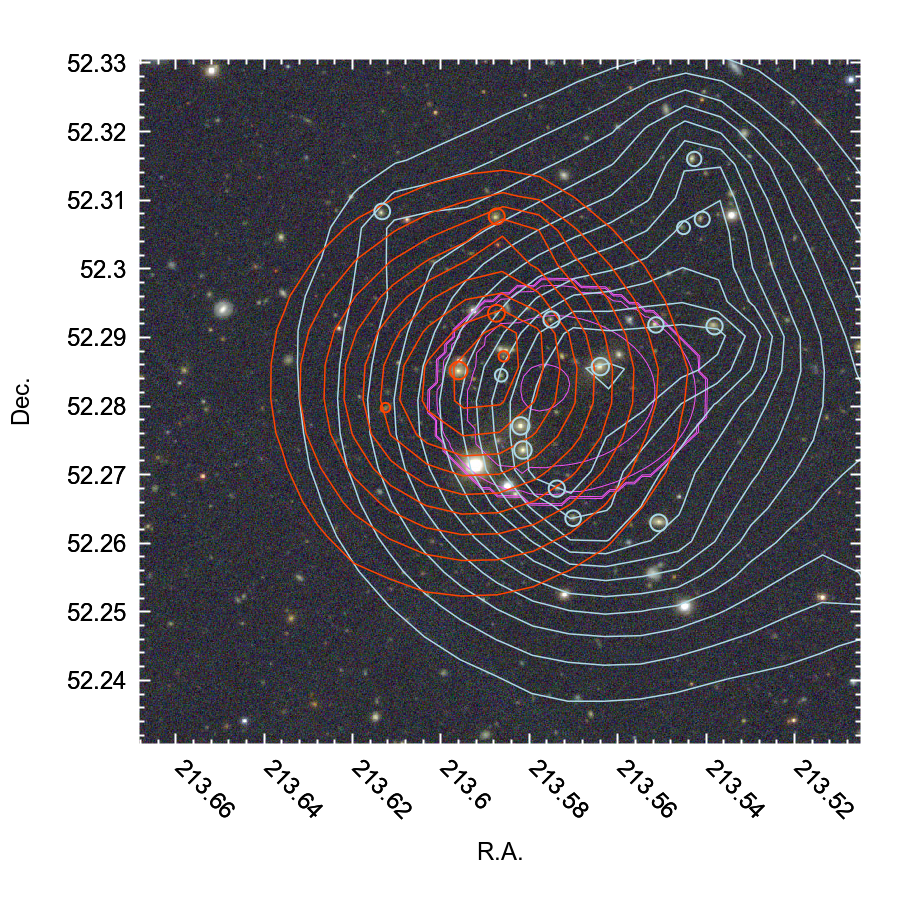}
    \includegraphics[width=0.52\textwidth]{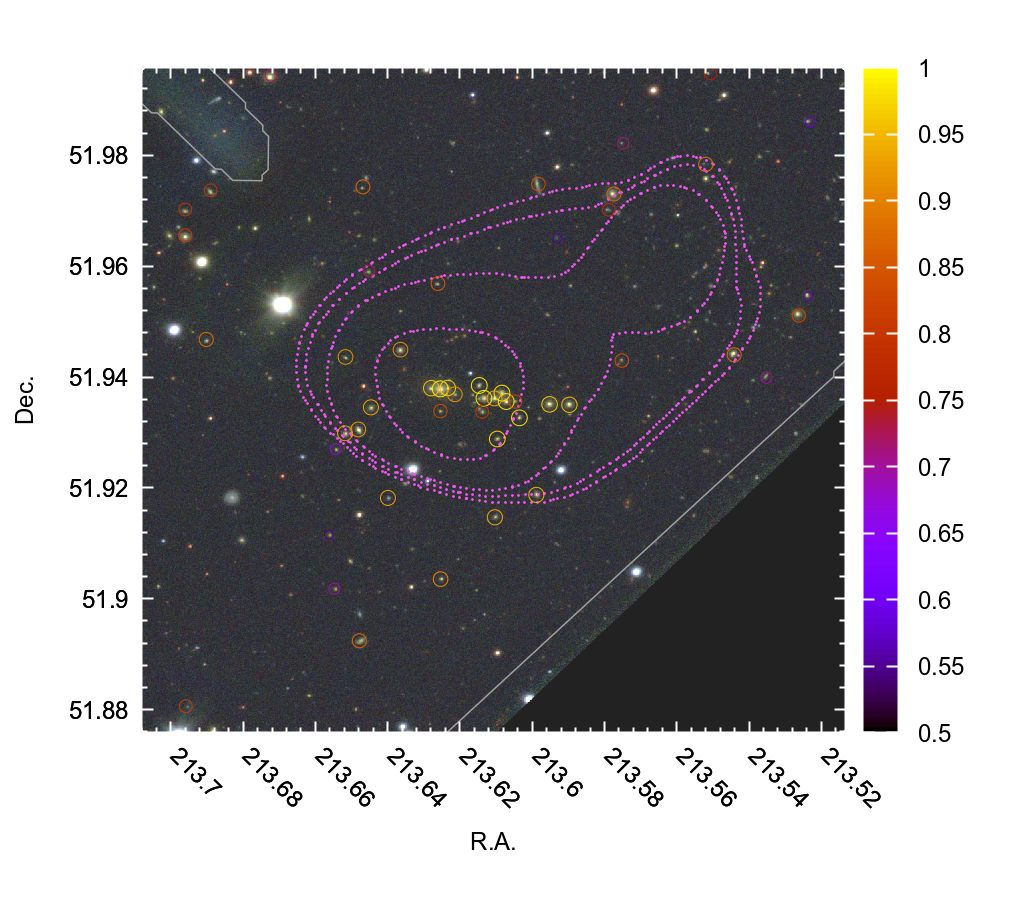}
    \caption{Left panel: two clusters located along the same line-of-sight but at two different redshifts ($z=0.233 \pm 0.005$ and $z=0.391 \pm 0.005$) displayed together with the map of the amplitude $A$ measured by AMICO (cyan and red isodensity contours, respectively); the galaxy members with a probabilistic association $P>0.5$ are indicated by circles. The filter response of AMICO is shown with the blue (for the lower redshift cluster) and red isocontours (for the higher redshift cluster). Right panel: the most massive cluster in the sample having an X-rays mass of $M=8.6\pm0.7 \times 10^{13} M_{\odot}$ and located at $z=0.29$. Here, the color of the circles indicate the probabilistic membership association of the members belonging to. The white lines represent the masked areas and the survey limit. On both panels, magenta contours show the extended X-ray emission.}
    \label{fig:amico-detections_picture}
\end{figure*}

 As a by-product, the algorithm provides an estimate of the fraction of masked effective area for each detection. Such fraction is defined as $f_j = 1 - B(z_j) / B_{eff}(\vec{\theta}_j,z_j)$, where $j$ refers to the detection, the background $B$ is evaluated as if no mask is present and $B_{eff}$ is evaluated over the area not covered by masks. We speak of an `effective' area because each unmasked pixel in the evaluation of $B$ is weighted by the amplitude of the cluster model. To illustrate how the filter highlights the density fluctuations, we show the resulting amplitude in Fig.~\ref{fig:amico-amplitude-map} for a redshift slice at $z=0.24$ in the \mj data. The complete footprint of the survey is shown together with the masked areas (white contours). In the bottom left and top right boxes we display the $g,r,i$ color composite postage-stamps based on the \mj images of the two most significant detections at that redshift. The violet contours refer to the extended X-ray emission detected  in deep Chandra data \citep{Erfanianfar13}. The top left box shows the AMICO response for a background slice at a higher redshift, $z=0.36$, where another cluster is detected nearly along the same line-of-sight of the main central structure visible in the background figure at $z=0.24$. The asterisks in the top left panel indicate the position of the cluster visible at redshift $z=0.24$. Note the absence in the AMICO maps of cross-contamination between different redshifts. Also the stamp image on the right hand side shows a case of chance alignment between two structures. More about this and another case will be discussed in \S\ref{sec:catalogue}.

The same formalism naturally provides a probabilistic association for each galaxy, labeled with the index $i$, to a specific detection, labeled with the index $j$,
\begin{equation}
  \label{eq:probability}
  P_i(j) = \tilde{P}_{f,i} \frac{A_j C(z_j;\vec{\theta}_i-\vec{\theta}_j,m_i)p_i(z_j)}
  {A_j C(z_j;\vec{\theta}_i-\vec{\theta}_j,m_i) p_i(z_j) + N(m_i,z_j)}  \;.
\end{equation}
Since clusters overlap in the data space, more than one cluster association can be assigned to a galaxy through an iterative approach, in which the $\tilde{P}_{f,i}= 1-\sum_k^{j-1} P_i(k)$ term accounts for the previous memberships assigned to the $i$-th galaxy. This probabilistic membership is relevant for (1) the characterization of galaxy populations in terms of environment \citep{rosa22,julio22}, (2) the iterative removal of the contribution of larger detections when searching for smaller structures and (3) the definition of two richness estimates $\lambda$ and $\lambda_*$, see the following Equations~(\ref{eq:apprichness}) and (\ref{eq:starrichness}).

\subsection{Additional mass proxies}\label{sec:proxy}

In addition to the amplitude, $A$, which is the direct output of the optimal filtering procedure as given by Eq.~(\ref{eq:amplitude}), the probabilistic membership association provided by AMICO allows to define two other mass proxies, which are directly related to the cluster richness. The first one, $\lambda$, is defined as the sum of the probabilistic membership association of all galaxies to the $j$-th detection,
\begin{equation}
  \label{eq:apprichness}
  \lambda_{j} = \sum_{i=1}^{N_{gal}} P_i(j) \;,
\end{equation}
and represents the number of visible galaxies belonging to a detection. Consequently it depends on the survey magnitude limit and it is thus redshift dependent. The second mass proxy, $\lambda_*$, follows the same definition of $\lambda$, but the sum runs only over the galaxies brighter than $m_*+1.5$, where $m_*$ represents the bright-end side cut-off of the Schechter luminosity function, and within the virial radius, $R_{200}$,
\begin{equation}
  \label{eq:starrichness}
  \lambda_{*j} = \sum_{i=1}^{N_{gal}} P_i(j) \quad\mbox{with}\quad
  \left\{
  \begin{array}{lr}
    m_i<m_*(z_j)+1.5 \\
    r_i(j) < R_{200}(z_j)
  \end{array}
  \right.\;.
\end{equation}
The radius $R_{200}$ and the magnitude $m_*$ are given by the model used to construct the filter (see Section~\ref{sec:amico}). This mass proxy is nearly redshift independent whenever $m_*+1.5$ is brighter than the survey limit, which for our case is true up to $z=0.56$. Mass-scaling relations for these mass proxies have been derived in \cite{bellagamba18b} and in \cite{lesci22} through the use of information coming from weak-gravitational lensing and clustering of galaxy clusters \citep{sereno15,marulli21,moresco21}.

\begin{figure*}
    \centering
    \includegraphics[width=0.48\textwidth]{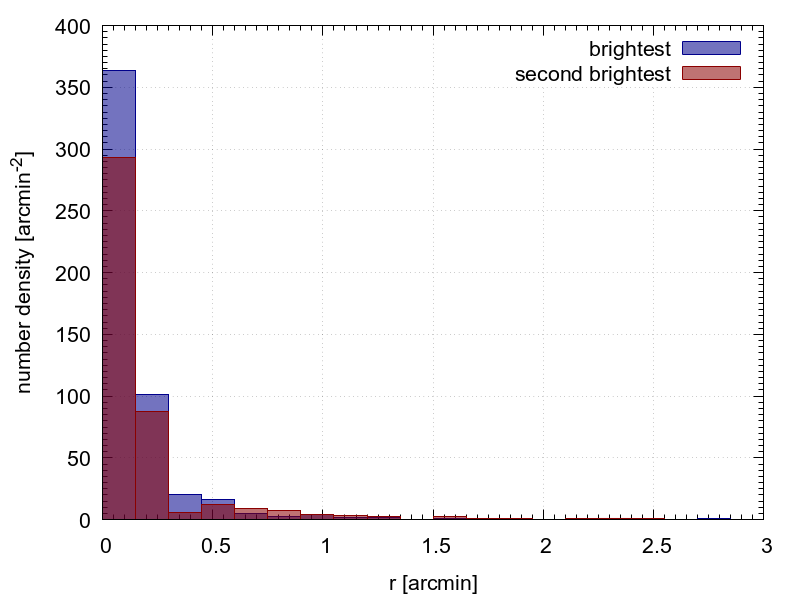}
    \includegraphics[width=0.48\textwidth]{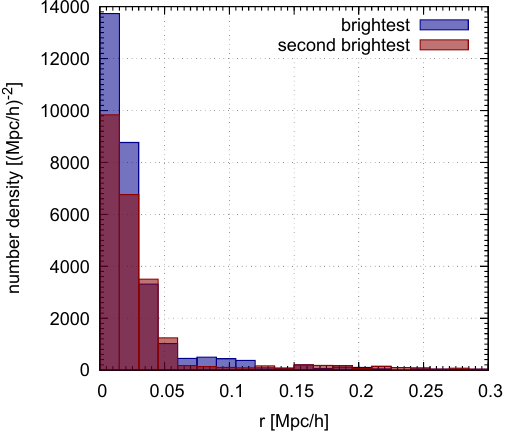}
    \caption{Distribution of the radial distance of the BGG candidates with respect to the cluster center identified by AMICO, given in arcmin (left panel) or Mpc/h (right panel). The BGGs are identified as the brightest and the second brightest galaxies with probabilities larger than 80\%.}
    \label{fig:CD-position}
\end{figure*}

The two richness estimates for all detections are displayed in the left and middle panels of Fig.~\ref{fig:amico-detections_proxies}. As it is clear from the discontinuity at $z=0.56$ where the $m_*+1.5$ limit is reached, the two richness estimates become strongly redshift dependent and their use as mass proxies requires a calibration to correct for this effect. This is not the case for the amplitude, $A$, because the filtering formalism automatically accounts and compensates for the survey magnitude limit and the redshift dependency can be safely neglected \citep[see also][]{maturi19}. This is why in this study we favour this mass proxy over $\lambda_*$, whose redshift dependency can not be properly calibrated because of the limited statistics offered by the $\sim1$ deg$^2$ area covered by the \mj data and the large cosmic variance. This limitation will be overcome when the wider area of teh J-PAS survey will be available.

In addition to the "standard" mass proxies of AMICO, in this work we defined a mass-proxy incorporating the stellar mass, $M_\star$, of individual galaxies \citep[see e.g.][]{pereira18}
\begin{equation}
  \label{eq:massstarrichness}
  \lambda_{M_{\star}j} = \sum_{i=1}^{N_{gal}} M_{\star j} P_i(j)
  \quad\mbox{with}\quad P_i(j)>0.5 \;.
\end{equation}
The usage of this quantity is possible thanks to the 56 J-PAS filters that enable us to perform a reliable Spectral Energy Distributions (SED) fitting for almost all galaxies in the sample, which has been done using the BaySEAGal code \citep{gonzalezdelgado21}, placing J-PAS in a quite unique position. It was in fact possible to evaluate $\lambda_{M_\star}$ for all groups but one located at $z=0.76$. 

In the following Section~\ref{sec:xrays} we will also obtain X-ray mass estimates \citep[based on weak lensing calibrations of X-ray luminosity]{Leauthaud10}. We will use them to define the mass-proxy scaling relations for the amplitude, $A$, for the intrinsic richness, $\lambda_*$, and for the total stellar mass, $\lambda_{M_\star}$, see Fig.~\ref{fig:scaling_all_Xrays}. More details are given in Section~\ref{sec:xrays-scaling}.

\subsection{The catalogue of galaxy clusters}\label{sec:catalogue}

In this work, we construct the filter adopting our cluster model, defined as $C(z_j; \vec{\theta}-\vec{\theta}_j, m) = R(z_j;|\vec{\theta}-\vec{\theta}_j|) \, L(z_j;m)$, i.e. the product of a Schechter luminosity function, $L$, in the $r$ band given by the combination of a passive and star forming populations of members with $\alpha_{red}=-0.53$, $\alpha_{blue}$ = -1.0, respectively, and a common $M_{*}=-20.8$. These values, taken from  \cite{hansen09},
refer to clusters with mass $M=10^{14} M_{\sun}$ and richness $N_{200}=25$ as estimated through the weak lensing mass-proxy scaling relation given by \cite{johnston07}. These values just serve as a guidance to define the template and do not need to be fine tuned. Same goes for the radial profile, $R$, describing the projected density distribution of cluster galaxies, here taken from the tabulated values given in \cite{sheldon09}.

The statistical properties of the noise, $N$, are extracted directly from the data under the assumption that the number of cluster members is small with respect to that of the overall population of galaxies. Such data driven approach suffers from the small area of \mj  because of the limited statistics and the results presented in this work are thus penalized in terms of sensitivity and contamination. Better performances will be certainly obtained, when the thousands of square degrees of J-PAS data will be available. Despite this limitation, the application of AMICO to \mj allows us to probe groups and clusters of galaxies over a wide range of redshifts and masses, as it will be shown below.

The cluster catalog provided by AMICO contains $80/30/11$ entries for signal-to-noise ratios (defined as $S/N=A/\sigma$ given by equations~\ref{eq:amplitude} and ~\ref{eq:amplitudesigma}) larger than $2.5/3.0/3.5$, respectively, and covers a mass range of approximately $10^{13}<M<10^{14} M_\odot/h$. Assuming this number density of clusters we expect to detect of the order of approximately $2.5\,\times10^5$ galaxy clusters with $S/N>3$, when all 8000~deg$^2$ of J-PAS will be observed. The redshift distribution of galaxies and the cluster candidates are shown on the left and central panels of Fig.~\ref{fig:amico-detections_z_distri}, respectively. The distribution of the detections signal-to-noise ratios is shown in the right panel of the same figure. The variations in both galaxy and cluster density are dominated by the physical fluctuations in the distribution of matter along the line-of-sight, see for example the "wall" at $z\approx0.28$.

In Fig.~\ref{fig:amico-detections_picture} we display some examples of detections. The left panel shows two groups with masses  $M_{200_c}=(6.5 \pm 0.5) 10^{13}\, M_\odot$ and $M_{200_c}=(6.1 \pm 0.5)\, 10^{13} M_\odot$ located at different redshifts, $z=0.228 \pm 0.005$ and $z=0.338 \pm 0.005$, but close in observer's plane. The detection at lower redshift is the one on the right hand side of Fig.~\ref{fig:amico-amplitude-map} (their mass has been derived thanks to X-ray observations as discussed in Section~\ref{sec:xrays}). Their amplitude maps are shown by the iso-density contours (red and cyan lines for the more and less distant detection, respectively), together with their members having a probabilistic association larger than $50$\%. These two objects are clearly disentangled by AMICO in contrast with the X-ray analysis in which only the combined flux can be measured (see the X-ray surface brightness displayed by the magenta contours in the right box of Fig.~\ref{fig:amico-amplitude-map}). This is a nice example in which X-ray mass measurements have to be discarded because of the mutual cross contamination of two cluster signals. This result will be also confirmed by the following analysis of the mass-proxies scaling relations: in fact, as evident from Fig.~\ref{fig:scaling_all_Xrays}, the X-ray mass for one of the two clusters, as well as for other cases of line-of-sight alignments, is largely overestimated. In such cases, removing optical counterparts susceptible to chance assignment to X-rays can improve the situation \citep{klein19}. The right panel of Fig.~\ref{fig:amico-detections_picture} shows the most massive object found in miniJPAS, a cluster with an X-ray mass of $M_{200_c}=(8.6 \pm 0.7) 10^{13}\, M_\odot$ located at redshift $z=0.290 \pm 0.005$. Here, the color of the circles quantifies the probabilistic membership derived with AMICO as indicated in the color-bar \citep[for a detailed discussion about this cluster and its members see][]{julio22}. Since the cluster model is based on a luminosity function and a radial density distribution of galaxies in clusters, the probability associated to each galaxy depends not only on the galaxy redshift, but also on their angular separation from the cluster center and on their magnitude: the closer to the cluster center and/or the brighter the galaxy is, the more likely is its membership to the cluster. The magnitude dependency is given by the difference in the luminosity functions of cluster members and field galaxies, where the first one has a more pronounced value at lower magnitudes.
The catalogue of galaxy cluster detections is available online\footnote{\link} and the quantities reported in its columns are described in Table~\ref{tab:catalogue_columns}.

\section{The catalogue of cluster members and BGGs}\label{sec:members-bcg}

\begin{figure}
    \centering
    \includegraphics[width=0.49\textwidth]{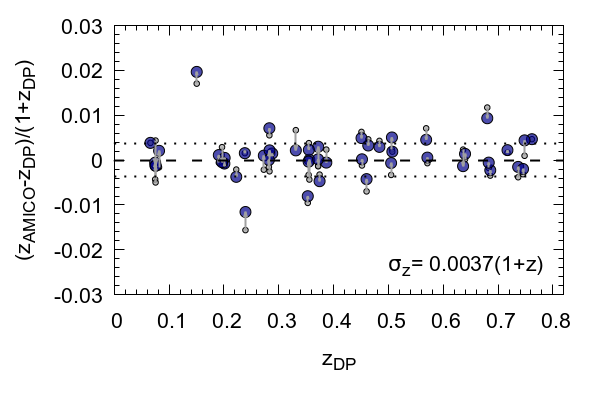}
    \caption{The redshift of the AMICO clusters based on the median \mj photo-$z$ of the galaxies with probabilistic association $P>0.5$ (blue circles) and within $10^\prime$ from the detection center as compared with the spectroscopic redshift based on DEEP3 data. The uncertainty of our redshifts estimate is of $\sigma_z=0.0037(1+z)$ (uncertainty drown with the horizontal dotted lines), which is approximately ten times better than what typically achieved with photometric surveys. The improvement of the redshift estimate based on the AMICO probabilistic memberships over the original output (small gray circles) is visible.}
    \label{fig:detections_z}
\end{figure}

\begin{table*}
    \centering
    \begin{tabular*}{\textwidth}{l@{\extracolsep{\fill}}llr}
        \hline
        \multicolumn{1}{l}{Column} & \multicolumn{1}{l}{Unit} & \multicolumn{1}{c}{Description} & \multicolumn{1}{r}{Example} \\
        \hline
    \texttt{ID} (1) & & identification number of the galaxy & 1 \\
    \texttt{NASSO} (1) & & number of clusters associated to the galaxy & 3 \\
    \texttt{IDASSO} (1) & & vector of size NASSO with the identification number of each associated cluster & 1, 1001, 3 \\
    \texttt{PASSO} (1) & & vector of size NASSO with the probabilistic association to each cluster & 0.34, 0.52, 0.02 \\
        \hline
    \end{tabular*}
    \caption{Descriptions of the columns for the AMICO probabilistic association of galaxies to clusters. The full VAC is available online at: \link}
    \label{tab2}
\end{table*}

With this catalogue of galaxy cluster members at hand, we created a sample of Brightest Group Galaxies (BGG) candidates by selecting the brightest galaxy member with a probabilistic membership larger than $80$\%. We also listed a second candidate by taking the second most luminous galaxy, again above the $80$\% probability threshold. This is done because the study of the magnitude gap between the first and second brightest galaxies is of interest \citep[e.g.][]{gozaliasl14}, for instance, it might be used to evaluate the presence of fossil groups. Our sample contains several small groups in which it is difficult to identify a clear dominant galaxy. For the most massive systems the presence of two bright central galaxies with a similar luminosity might indicate a recent major merger event where the two galaxies might have been the BGGs of the parent structures. In Fig.~\ref{fig:CD-position} we plot the radial distance between the so identified BGG and the cluster center as defined by AMICO. Clearly, the second brightest galaxies (BGG with rank 2) are on average more distant from the cluster center with respect to the most brightest galaxy (BGG with rank 1), as expected.

\section{Spectroscopic analysis: redshift of clusters and probabilistic membership accuracy}\label{sec:spectral}

The DEEP3 galaxy redshift survey partially overlaps the footprint of the miniJPAS survey \citep{bonoli20}. We can then use this data set to derive a spectroscopic redshift estimate for the AMICO clusters within the DEEP3 region and to select their members. We used these estimates to test the redshift and membership association provided by AMICO.

\begin{figure}
    \centering
    \includegraphics[width=0.45\textwidth]{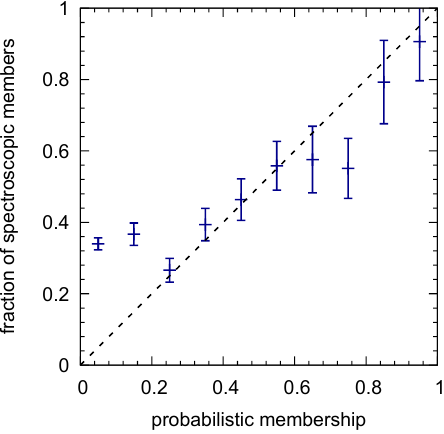}
    \caption{Probabilistic membership association of galaxies to clusters returned by AMICO compared with the fraction of spectroscopic cluster members identified with the `shifting gapper' technique applied to the DEEP3 data. The photometric probabilistic membership is in good agreement with the spectroscopic one except for probabilities smaller than $P<0.2$.}
    \label{fig:probabiliti-spetral}
\end{figure}

We started by deriving a list of spectroscopic cluster members using the `shifting gapper' technique in radial bins from the cluster centre \citep{fadda96} as described in \citet{lop09,lop14}. 
The method is based on the application of the gap-technique in radial bins (described in \citealt{kat96}), to identify gaps in the redshift distribution. Instead of adopting a fixed gap, such as 1000 $km$ $s^{-1}$ or $0.005(1+z)$, we considered a variable gap, called {\it density gap} \citep{ada98, lop09}. The density gap size is given by the expression $\Delta z = f_g(1+\exp(-(N-6)/33))/c$, where $N$ is the number of galaxies found in the redshift survey of a cluster \citep{ada98}, and $c$ is the speed of light in $km/s$. Note that the gap parameter $f_g$ replaces the fixed value of ``500'' adopted in \citet{ada98}. This gap factor, $f_g$, scales with the velocity range of the galaxies found in each radial bin and is a better choice to work with systems of different masses \citep{lop09}. We considered all galaxies within 2.5 Mpc/h (3.57 Mpc for h $= 0.7$) from the cluster center and with $c\, |z - z_{cl}| \le 4000~{\rm km~s}^{-1}$, where $c$ is the speed of light and $z_{cl}$ the cluster redshift.  We also used a bin size of 0.42 Mpc/h (0.60 Mpc for h $=0.7$) or larger if less than 15 galaxies have been selected. In every radial bin we discarded galaxies not associated to the main body of the cluster (those with a velocity difference exceeding the velocity gap). 
The procedure is repeated until no more galaxies are rejected as interlopers. One great advantage of this method is to make no hypothesis about the dynamical status of clusters that do not need to be virialized. In the end we were able to obtain spectroscopic redshift estimates and spectroscopic memberships for 47 AMICO clusters. Given the relatively small cluster masses and the limited number of members per cluster with spectroscopic redshift it was not possible to derive reliable estimates of clusters properties such as $\sigma_{cl}$, $R_{200}$ and $M_{200}$. The comparison with the redshift returned by AMICO is shown in Fig.~\ref{fig:detections_z} (gray circles): the scatter is very small $\sigma_z=0.0053(1+z)$, dominated by the redshift sampling used in the run rather than by the photo-$z$s uncertainty, and no bias is present.

\begin{figure*}
    \centering
    \includegraphics[width=0.45\textwidth]{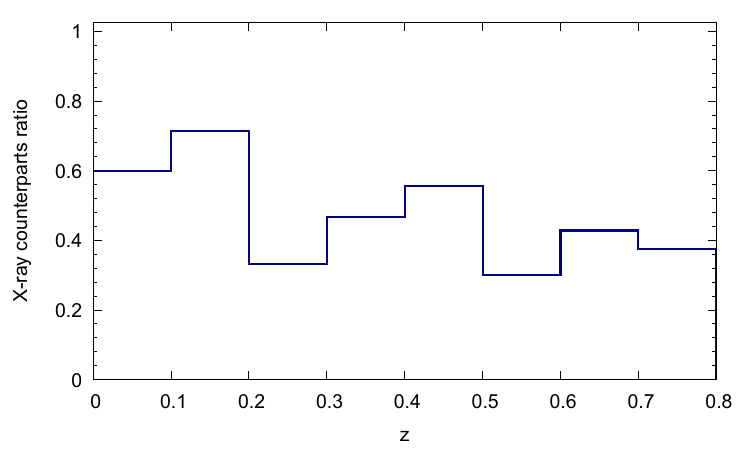}
    \includegraphics[width=0.45\textwidth]{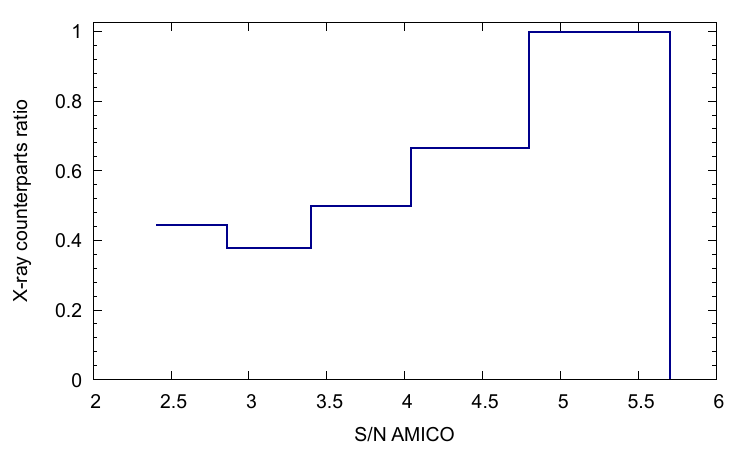}
    \includegraphics[width=0.45\textwidth]{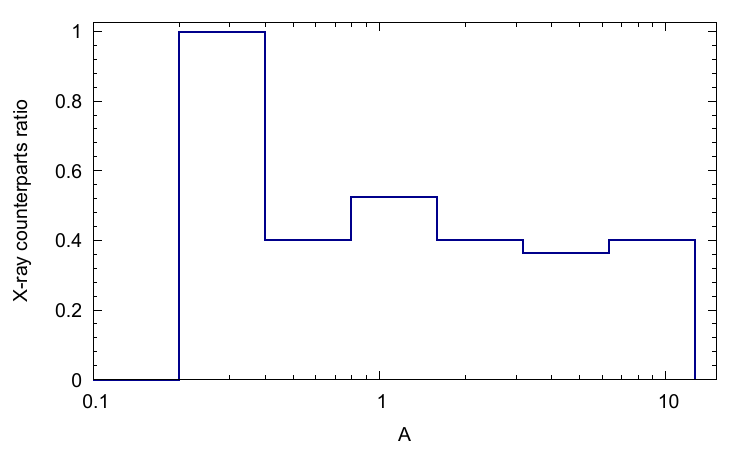}    \includegraphics[width=0.45\textwidth]{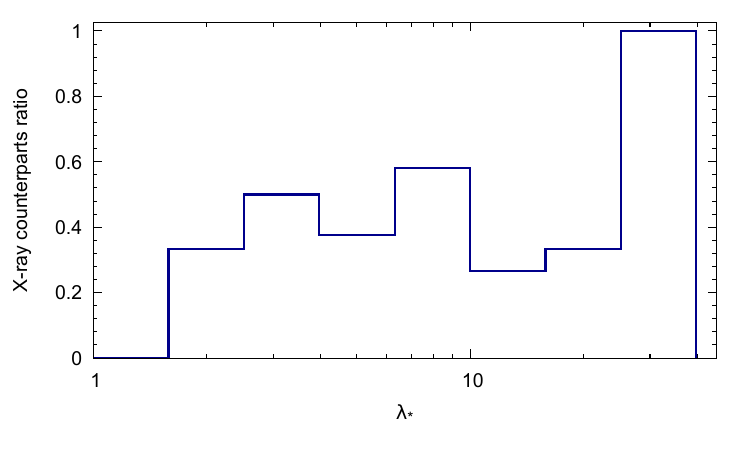}
    \caption{Fraction of optical detections with an X-ray counterpart with $S/N>1$ as a function of redshift (top-left panel), AMICO signal-to-noise ratio (top-right panel), amplitude $A$ (bottom-left panel) and intrinsic richness $\lambda_*$ (bottom-right panel).}
    \label{fig:xrays-matches}
\end{figure*}

To refine the results based on the photometry only and the membership provided by AMICO, we estimated the cluster redshifts as the median of the redshifts of the galaxies with an AMICO probabilistic membership association larger than $P_i(j)>0.5$ and within $10^\prime$ from the detection center (see the blue circles in Fig.~\ref{fig:detections_z}). This approach provides a smaller redshift uncertainty, $\sigma_z=0.0037(1+z)$, 
with respect to the one directly obtained with AMICO (gray points) which suffers from the relatively coarse redshift resolution, $\Delta z=0.01$, used when running the algorithm. The procedure seems to be stable as the redshift correction is at most half `redshift pixel', as expected. A higher resolution would be demanded by the high accuracy of the miniJPAS photo-$z$s, but this is not possible since the present limited area of the survey does not allow a reliable noise estimate in more narrow redshift slices, as already discussed. A redshift pixellization with an higher resolution will be adopted for the J-PAS data. Note that the resulting redshift uncertainty is about $10$ times better than what is typically obtained with broad band filters \citep[see e.g.][]{maturi19}.

This high redshift accuracy makes J-PAS an ideal survey to derive cosmological constraints based on the clustering of galaxy clusters. When performing such studies, the redshift uncertainty introduces an exponential suppression on the redshift-space 2D power spectrum causing a scale-dependent removal of signal over a typical scale $k \sim \sigma^{-1}$, where $k$ is the modulus of the wave-vector components parallel and perpendicular to the line-of-sight and
\begin{equation}
    \sigma \equiv \frac{c\sigma_z}{H(z_m)} \;.
\end{equation}
Here $H(z_m)$ is the Hubble function computed at the median redshift of the cluster sample, $z_m$, and $\sigma_z$ is the typical photo-$z$ error \citep[see e.g.][]{sereno15,lesci22}.
Hence, gaining a factor of 10 in the photo-$z$ accuracy with respect to broad band filter photometric surveys will allow us to have a stronger clustering signal at scales 10 times smaller then what achievable with broad band photometric surveys \citep{veropalumbo14}.

We then used the cluster memberships derived using the spectroscopic information to test the AMICO probabilistic membership association given in Eq.~(\ref{eq:probability}). The comparison between the fraction of spectroscopic members with the probabilistic memberships association provided by AMICO is shown in Fig.~\ref{fig:probabiliti-spetral}. 

\begin{figure*}
    \centering
    \includegraphics[width=0.42\textwidth]{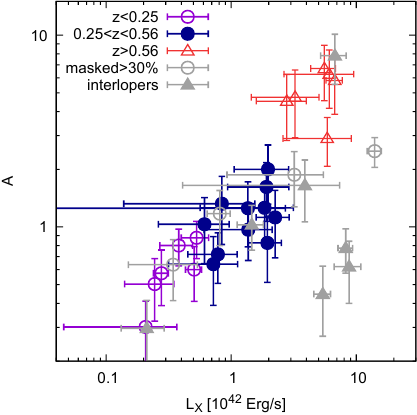}
    \hspace{0.5cm}
    \includegraphics[width=0.42\textwidth]{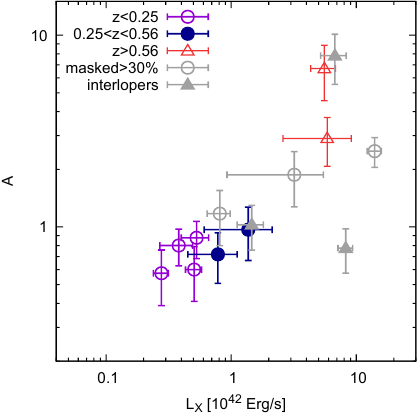}
    \caption{Relation between the AMICO amplitude, $A$, and the X-ray luminosity on which the X-ray mass estimates are based (albeit with a redshift-dependent correction). The gray points have been rejected from this analysis because they are either heavily masked (more than $30$\% of the cluster) or their X-ray flux is possibly contaminated by an interloper. Only the systems with a significance of X-ray flux estimate above the $1\sigma$ significance are shown. Left (right) panel refers to optical detections with
    $S/N > 2.5$ ($S/N > 3.0$).}
    \label{fig:A-Lx}
\end{figure*}

\begin{figure*}
    \centering
    \includegraphics[width=0.42\textwidth]{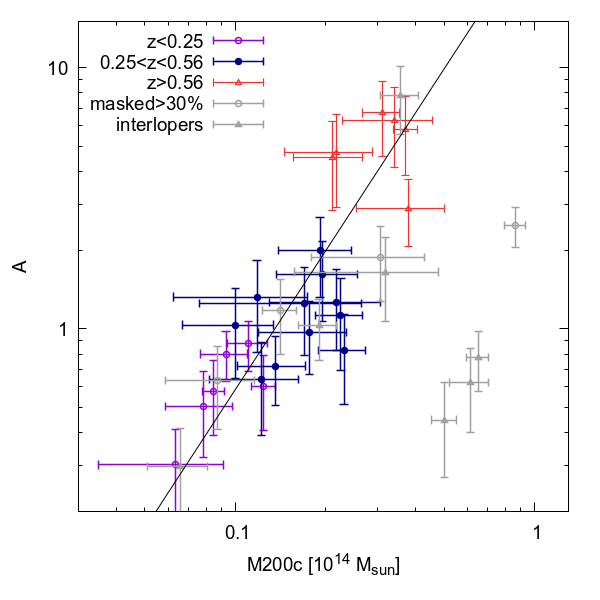}
    \hspace{0.5cm}
    \includegraphics[width=0.42\textwidth]{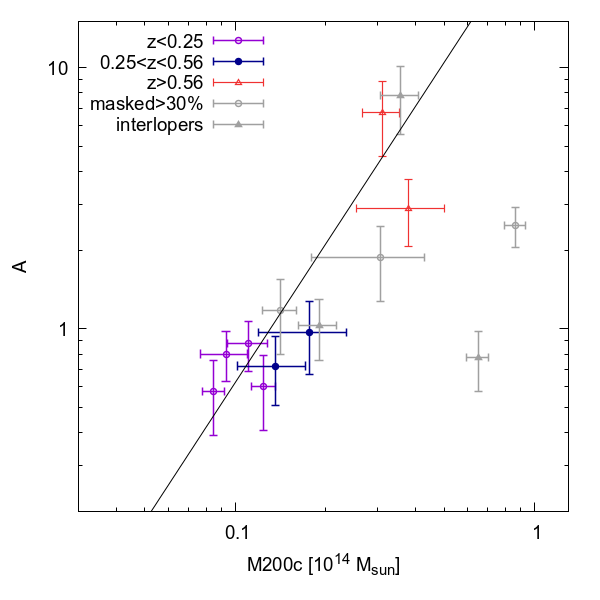}
    \includegraphics[width=0.42\textwidth]{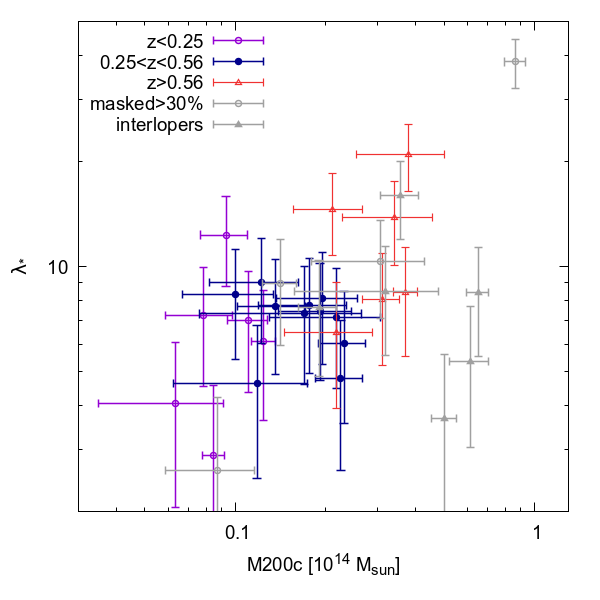}
    \hspace{0.5cm}
    \includegraphics[width=0.42\textwidth]{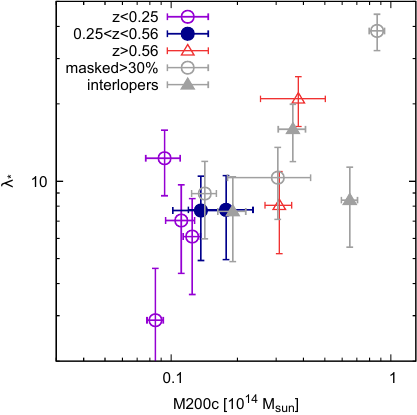}
    \includegraphics[width=0.42\textwidth]{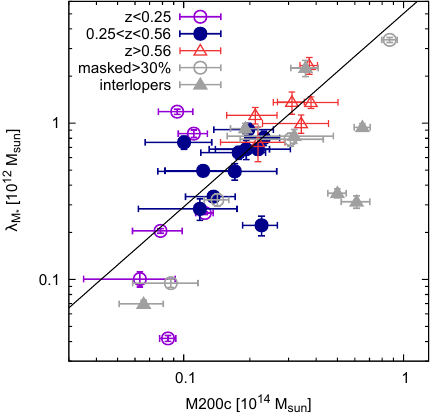}
    \hspace{0.5cm}
    \includegraphics[width=0.42\textwidth]{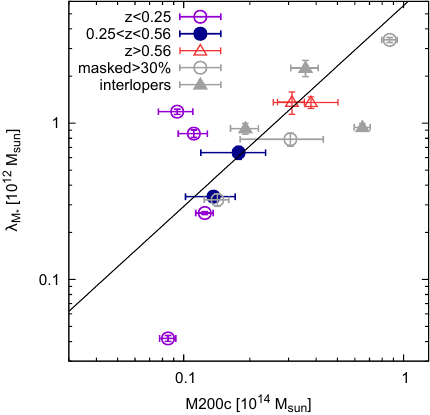}
    \caption{Scaling relations between different mass proxies ($A$, top panels, $\lambda_*$, central panels, and $\lambda_{M_\star}$, bottom panels) and our X-ray mass estimates for the cluster sample with X-ray fluxes above the X-ray $1\sigma$ limit. Results are shown for two different optical significance: $S/N> 2.5$ and $S/N> 3.0$, in the left and right panels, respectively. The best-fit scaling relations have been reliably derived for $A$ and $\lambda_{M_\star}$ but not for $\lambda_*$. For the fit we used only those detections with an optical mask fraction smaller than $30$\%, with no X-ray contamination given by interlopers identified with AMICO.}
    \label{fig:scaling_all_Xrays}
\end{figure*}
For the most interesting interval of probabilities, i.e. $P>0.2$, the AMICO probabilistic associations appear in very good agreement with the spectroscopic memberships. For lower values, $P<0.2$, there is a small discrepancy between spectroscopy and the result provided by AMICO. This might be due to the spectroscopic memberships (not due to AMICO) or to the cluster model used by AMICO during the detection phase, where the scale radius is larger than the typical one for the systems in the sample. While we plan to investigate this issue in more details in future work, its importance is marginal, since the galaxies in question are well below the 50\% membership cut in all estimates, live in the outskirt regions, and, in general, are not used in follow-up studies. The effect of this bias on the $\lambda_*$ and $\lambda_{M_\star}$ estimates, which are based on the membership probabilities, is not going to be that relevant because it affects the galaxies with the lowest probabilistic memberships which contribute the least and which tend to be faint galaxies, likely excluded by the magnitude cut-off embedded in the definition of these mass-proxies.

\section{Analysis of the cluster sample using X-ray data}\label{sec:xrays}
\subsection{X-ray data}\label{subs:xray}
The \mj data comprise the area covered by the All-Wavelength Extended Groth Strip International Survey \citep[AEGIS]{davis07}. The X-ray data, based on several Chandra and XMM-Newton campaigns,  result in one of the deepest X-ray cluster catalogs on the sky \citep{Erfanianfar13}.

The \mj data cover a larger area compared to the X-ray footprint analyzed in \cite{Erfanianfar13}, but the area outside of it has been extensively observed by XMM-Newton (PI A. Merloni). We have reanalyzed all XMM-Newton observations overlapping with the \mj data using the latest XMMSAS (version 18.0.0) and following the procedures on flare screening, background estimate, point source removal and image mosaicing outlined in  \citet{Finoguenov10} and \citet{Erfanianfar13}. To measure the X-ray flux of the AMICO clusters, we combined point source cleaned Chandra and XMM-Newton data and placed a $0.6^\prime$ aperture on the central position of each AMICO source. The relatively large aperture is meant to cope with possible miscentering  between optical and X-ray positions. This strategy reduces possible biases in the estimates of the X-ray fluxes with the cost of a slightly larger uncertainty on the derived quantities. A subsequent computation of the X-ray luminosity iteratively takes into account the fraction of the flux retained within the aperture, and computes the K-correction, based on the redshift of the source and its spectral shape governed by the temperature estimated using  an $L-T$ relation \citep{Finoguenov10}. The weak lensing calibration of the relation between X-ray luminosity and the total mass ($M_{200c}$) has been derived using similarly derived luminosities \citep{Leauthaud10,taylor12}, so any biases of the procedure are absorbed by the calibration. The cosmological parameters adopted for this weak lensing calibration are the same assumed in the cluster detection procedure.

For the non-detected clusters, we place the $1\sigma$ upper limit on the X-ray properties. The fraction of optical detections with an X-ray counterpart as a function of redshifts, optical (AMICO) signal-to-noise ratio, amplitude $A$, and  richness $\lambda_*$ is shown in Fig.~\ref{fig:xrays-matches}. No particular trend is visible except for a clear increase of such fraction towards detections with higher signal-to-noise ratios which are more likely to be true positives associated to lower redshift and more massive systems, whose X-ray emission is more likely to be detected. It also seems that there is a difference in the evolution of X-ray vs AMICO efficiency to detect clusters, with X-rays depths yielding to AMICO at $z>0.5$. This highlights that in order to fully calibrate the high-$z$ performance of AMICO on J-PAS, much deeper X-ray data are needed, which will only be readily available with future X-ray missions, such as {\it Athena}. 

In Fig.~\ref{fig:A-Lx}, we show the relation between the AMICO amplitude, $A$, and the X-ray luminosity. A strong correlation between these two quantities shows how optical proxies can be used to derive reliable mass scaling relations, as it will be discussed in Section~\ref{sec:xrays-scaling}. From this data set, it appears that clusters down to very small amplitudes (X-ray flux) have a very "well-behaved" relation between their ratio of dark matter, baryonic content (both in terms of gas and number of members), and X-ray fluxes. Only four outliers stand out in the plot, but their X-ray flux is unreliable due to interlopers, i.e. objects belonging to pairs or triplets of clusters displaced along the same line-of-sight within $r<1^\prime$,  (three of them, see for instance left panel of Figure~\ref{fig:amico-detections_picture}) or because of an underestimation of the AMICO amplitude due to the non-complete optical coverage of the cluster, more than $30$\% of its area falls outside the \mj footprint (see right panel of Figure~\ref{fig:amico-detections_picture}).

\subsection{Mass-proxy scaling relations}\label{sec:xrays-scaling}

We model the relation between our X-ray ($M_{200c}$, obtained for each AMICO cluster following the procedures described in \S\ref{subs:xray}) and AMICO ($O$) mass proxies, discussed in Section~(\ref{sec:proxy}) as
\begin{equation}\label{eq:proxyScaling}
    log_{10}\frac{M_{200c}}{10^{14}M_\odot/h} = \alpha + \beta\,\log_{10}\frac{O}{O_{piv}} \,,
\end{equation}
where $O_{piv}$ is a pivot value \citep{bellagamba18b}. We do not account for any possible redshift dependency because the current sample is too small for its solid assessment, but it will be included when analysing future data releases. The best fit relations for $A$ and $\lambda_{M_\star}$ are shown in Fig.~\ref{fig:scaling_all_Xrays} for detections with $S/N>2.5$ and $S/N>3.0$ (left and right panels, respectively). No scaling relation was derived for $\lambda_*$ because of the excessive scatter and strong redshift dependence.

\begin{figure*}
    \centering
    \includegraphics[width=0.49\textwidth]{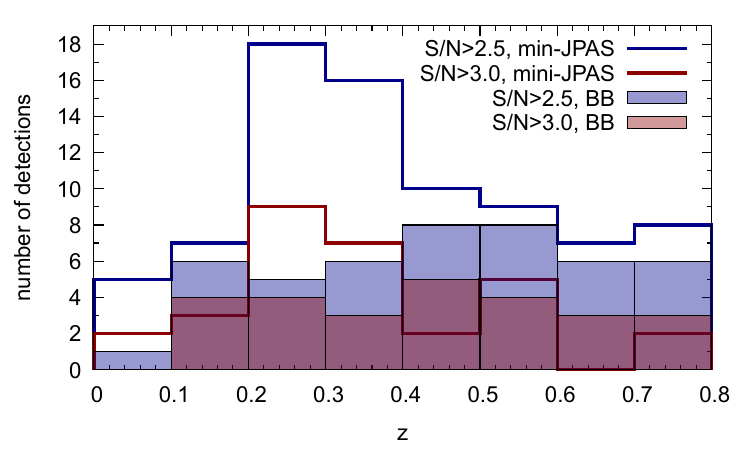}
    \includegraphics[width=0.49\textwidth]{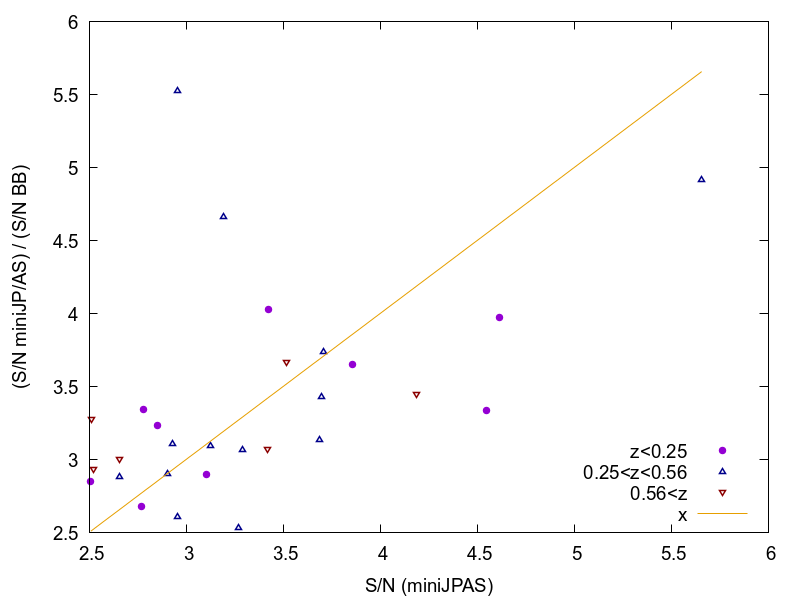}
    \includegraphics[width=0.49\textwidth]{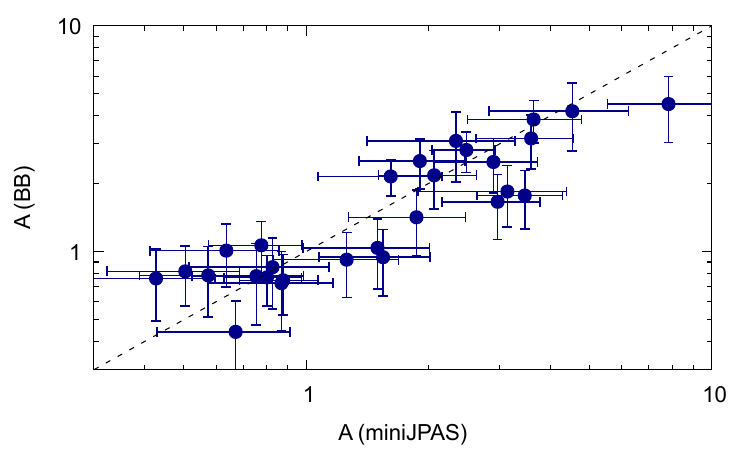}
    \includegraphics[width=0.49\textwidth]{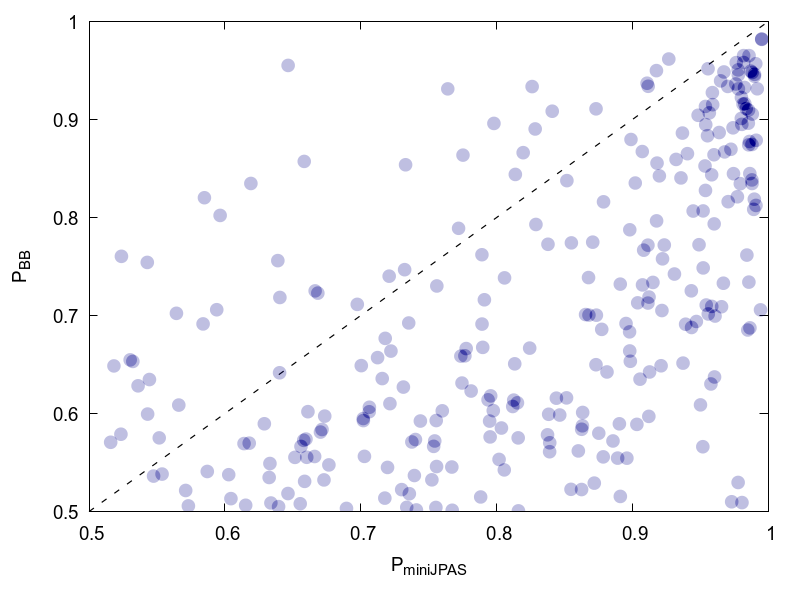}
    \caption{Comparison between the properties of the detections derived from the catalogue of galaxies with degraded photo-$z$ mimicking a typical photometric survey with redshift uncertainty $\sigma_z=0.03(1+z)$ (broad-band, BB), and of the detections based on the original miniJPAS data exploiting the 56 J-PAS filters. We show in the top left panel the redshift distribution of the detections (solid boxes refer to the degraded broad-band like data and the lines to the original data set; detections with $S/N>2.5$ and $S/N>3$ are shown in blue and red, respectively), in the top right panel the signal-to-noise ratios, in the bottom left panel the amplitude $A$ and in the bottom right panel the probabilistic membership association of galaxies to clusters as provided by AMICO.}
    \label{fig:degraded-SN}
\end{figure*}

Despite the fact that only four clusters are real outliers, when fitting the scaling laws, we opted to use a hard criteria for the sample selection, i.e. we considered only the clusters with an X-ray flux significance above the $1\sigma$ limit and excluded eleven clusters with either significant optical masking or line-of-sight projection affecting X-ray flux estimate. Four rejected clusters, indicated with open gray circles, are heavily masked (more than $30$\% of the cluster). Even if AMICO corrects for the masked fraction, one of them remains a clear outlier. The other seven, indicated with gray solid triangles, are systems affected by interlopers.
The values of the best fit parameters based on the mass-proxy scaling relations are reported in Table~\ref{tab:proxyBstFit}. We only list the results for two mass proxies, namely $A$ and $\lambda_{M_\star}$, because of the excessive scatter in $\lambda_*$ and because the scaling relation is currently very sensitive to the signal-to-noise ratio cutoff adopted for the cluster sample.
\begin{table}
    \centering
    \begin{tabular*}{0.48\textwidth}{llll}
        \hline
        Proxy & $\alpha$ & $\beta$ & $P_{piv}$ \\
        \hline
        $A$ (S/N>2.5) & $-0.69 \pm 0.03$ & $0.57 \pm 0.07$ & $2.0$\\
        $A$ (S/N>3.0) & $-0.71 \pm 0.05$ & $0.57 \pm 0.12$ & $2.0$\\
        $\lambda_{M_\star} (S/N>2.5)$ & $1.05 \pm 0.43$ & $0.80 \pm 0.19$ & $10^{12}$ \\
        $\lambda_{M_\star} (S/N>3.0)$ & $0.97 \pm 0.99$ & $0.78 \pm 0.43$ & $10^{12}$ \\
        \hline
    \end{tabular*}
    \caption{Best fit values for the mass-proxy scaling relations given by Equation~(\ref{eq:proxyScaling}) for the sample with $S/N>2.5$ and $S/N>3.0$. We do not show the scaling relations for $\lambda_*$ because of the excessive error on the values of the estimated parameters and the strong redshift dependence.}
    \label{tab:proxyBstFit}
\end{table}
For the current release we prefer to use the amplitude, $A$, as mass proxy because the best fit parameters of the scaling relations are very stable with respect to the choice of the signal-to-noise ratio adopted to select the clusters. Moreover, the filtering procedure provided by AMICO automatically accounts for the magnitude limit of the survey and the estimate appears unbiased over the entire redshift range under consideration. In contrast, the intrinsic richness, $\lambda_*$, and $\lambda_{M_\star}$ suffer from the galaxy sample incompleteness for redshifts $z>0.56$ as discussed in Section~(\ref{sec:proxy}). When it will be possible to constrain their redshift dependency, also $\lambda_*$ and $\lambda_{M_\star}$ will be valuable options even though they show a larger scatter around the best fit with respect to the amplitude, $A$. This might be because here we are dealing with very small system populated by few bright galaxies. We will investigate in depth this possible issue once data covering a larger area will be available. Finally, $\lambda_{M_\star}$ appears to be very promising as its scatter around the best fit relation is remarkably smaller in comparison to the one of $\lambda_*$, it is very sensitive to the mass and has the additional benefit of providing a physical characterization of clusters in terms of their stellar mass. This is one of the unique features of J-PAS, due to the excellent SED modelling it enables. 

We would like to stress that for this field we have very deep X-ray data (with a flux limit ranging from $10^{-15}$ to $3 \times 10^{-15}$ erg/s/cm$^2$ at $1\sigma$ level) which allows us to derive mass-proxy scaling relations down to very small masses, notably for individual clusters and not only through stacking. Such deep X-ray data sets are not going to be available for the large number of galaxy groups that J-PAS will provide in the upcoming future. For instance the ROSAT all-sky survey has a flux limit of $10^{-13}$/$10^{-12}$ erg/s/cm$^2$ for the deep/faint areas and the eRosita survey of $10^{-14}$ erg/s/cm$^2$ \citep{merloni12}. This poses the problem on how to improve the scaling relations for the smaller systems, which J-PAS can reach thanks to its sensitivity. This calls for possible targeted X-ray deep observations of the J-PAS galaxy groups, as well as the analysis of archival XMM-Newton data overlapping the J-PAS survey footprint. Weak-gravitational lensing and velocity dispersion measurements will also provide a way out, with large data sets already available within the J-PAS survey footprint from SDSS \citep[e.g.]{Kirkpatrick21}, CODEX \citep{Kiiveri21}, DECaLS \citep[e.g.]{Phriksee20} and in future from Euclid \citep{scaramella22} and DESI \citep{desi19}.

\section{Impact of narrow band photometry on galaxy cluster detection}\label{sec:impact-narrow-filter}

To quantify the gain in the cluster detection efficiency provided by the high quality of the photo-$z$s based on the 56 J-PAS filters, we degraded the data by introducing a final scatter in the photo-$z$ of $\sigma_z=0.03(1+z)$, to mimic the performances of a typical broad band survey \citep[see for instance][]{maturi19}. This assumption is optimistic in terms of scatter and, even more importantly, it does not account for possible biases that would affect such surveys. We recall that photo-$z$s based solely on broad band filters are more prone to biases caused by features in the galaxy SED, that they can not resolve, such as for example the $4000\AA$ break when transitioning between the $g$ and $r$ filters \citep[see e.g.][]{Padmanabhan05,maturi19,bonoli20,gonzalezdelgado21}. By applying AMICO to this degraded data set we detect only $46/26/9$ clusters, for signal-to-noise ratios larger than $2.5/3.0/3.5$ respectively, against the $80/30/11$ detections in the original catalogue. The increase in sensitivity given by the narrow band filter is remarkable, as shown by  Fig.~\ref{fig:degraded-SN}, where we plot the redshift distribution (top left panel) and compare the detection signal-to-noise of the original and degraded samples (top right panel). The gain in sensitivity that narrow band filters provide is clear. The amplitude, $A$, is only minimally affected showing the stability of the algorithm (bottom left panel), shown the stability of the algorithm. Finally, the probabilistic membership of galaxies to clusters is largely improved by the narrow band filters as expected (bottom right panel). To prove thi slast point numerical simulations are needed but here it is clear that galaxies with relatively high probabilistic membership ($P(z)>0.8$) in the BB survey have extremely high probabilities in the \mj data ($P(z)\sim0.98$.

\section{Conclusions}\label{sec:conclusions}

We applied the algorithm AMICO for cluster detection to the $\sim 1$ deg$^2$ area of the \mj data detecting $80$, $30$ and $11$ systems with signal-to-noise ratio larger than $2.5$, $3.0$ and $3.5$, respectively, in the redshift range $0.05<z<0.8$ down to a mass of $\sim 10^{13}\,M_\odot/h$. With this number density of galaxy clusters we derived, we expect to detect on the order of $2\,\times10^5$ galaxy clusters with $S/N>3$ in the upcoming 8000 deg$^2$ of J-PAS, with unprecedented sensitivity in mass and redshift accuracy for a photometric survey.

We derived mass-proxy scaling relations for the native response of AMICO, i.e. the so called amplitude $A$, and a newly defined estimate of the stellar mass $\lambda_{M_\star}$. The mass estimates for the clusters used for this analysis have been obtained  using the deep Chandra and XMM data available for the AEGIS field. The best fit values of the scaling relations parameters are listed in Table~\ref{tab:proxyBstFit}. We found that the amplitude $A$ has a smaller scatter around the best fit relation and is more robust with respect to the redshift dependency as compared to the other mass proxies because, by directly relying on the filtering formalism, it automatically accounts and compensates for the survey depth limit. In contrast, $\lambda_*$ and $\lambda_{M_\star}$ assume a fixed maximum absolute magnitude cut-off, which, for our data set, can not be reached for redshifts larger than $z=0.56$. Thus, they require a posteriori calibration which will be possible only when more data (i.e. larger area) will be available.

Furthermore, we produced a catalogue of cluster members based on the probabilistic association of galaxies to clusters provided by AMICO itself. We compared our probabilistic memberships with the one based on the "shifting gapper" approach applied to the spectroscopic DEEP3 survey. Our probabilistic membership appears to be in good agreement for galaxies with probability $P>0.2$, with differences observed at lower probabilities, which are anyway of lower interest being object likely not belonging to clusters. 
Such catalogue of members can be exploited  to study galaxy populations and to produce mock catalogues for the estimation of the purity and completeness of the cluster sample with data driven approaches such as the one implemented in SinFoniA \citep{maturi19}. These are crucial information needed for the derivation of cosmological constraints based on the cluster samples. Finally we used the memberships to identify the BGGs belonging to our cluster sample. We further show how the narrow band filters of J-PAS provide a substantial gain in sensitivity and an uncertainty on the redshift of clusters of only $\sigma_z=0.0037(1+z)$ placing J-PAS in between photometric and spectroscopic surveys. As an outlook, this might open up the possibility to study the clustering of galaxy clusters as well.

The performances of AMICO and J-PAS we demonstrate in this work will allow us to characterize galaxy-groups and -clusters {down to small groups of $\sim 10^3 M_\odot/h$}, identify the BGGs, split the galaxy population according to the environment and derive cosmological constraints. All data products are available online\footnote{\link}, together with the \mj data.

\begin{acknowledgements} 

Based on observations made with the JST250 telescope and PathFinder camera for the miniJPAS project at the Observatorio Astrof\'{\i}sico de Javalambre (OAJ), in Teruel, owned, managed, and operated by the Centro de Estudios de F\'{\i}sica del  Cosmos de Arag\'on (CEFCA). We acknowledge the OAJ Data Processing and Archiving Unit (UPAD) for reducing and calibrating the OAJ data used in this work.

Funding for OAJ, UPAD, and CEFCA has been provided by the Governments of Spain and Arag\'on through the Fondo de Inversiones de Teruel; the Arag\'on Government through the Research Groups E96, E103, E16\_17R, and E16\_20R; the Spanish Ministry of Science, Innovation and Universities (MCIU/AEI/FEDER, UE) with grant PGC2018-097585-B-C21; the Spanish Ministry of Economy and Competitiveness (MINECO/FEDER, UE) under AYA2015-66211-C2-1-P, AYA2015-66211-C2-2, AYA2012-30789, and ICTS-2009-14; and European FEDER funding (FCDD10-4E-867, FCDD13-4E-2685).

R.G.D. acknowledge financial support from the State Agency for Research of the Spanish MCIU through the "Center of Excellence Severo Ochoa" award to the Instituto de Astrof\'\i sica de Andaluc\'\i a (SEV-2017-0709), and through PID2019-109067-GB100.

R.A.D. acknowledges support from the Conselho Nacional de Desenvolvimento Científico e Tecnológico - CNPq through BP grant 308105/2018-4, and the Financiadora de Estudos e Projetos - FINEP grants REF. 1217/13 - 01.13.0279.00 and REF 0859/10 - 01.10.0663.00 and also FAPERJ PRONEX grant E-26/110.566/2010 for hardware funding support for the JPAS
project through the National Observatory of Brazil and Centro Brasileiro de Pesquisas Físicas. LM acknowledges support from the grants PRIN-MIUR 2017 WSCC32 and ASI n.2018-23-HH.0.

J.M.D. acknowledges the support of projects PGC2018-101814-B-100 and MDM-2017-0765.

VM thanks CNPq (Brazil) and FAPES (Brazil) for partial financial support.

AZ acknowledges support by Grant No. 2020750 from the United States-Israel Binational Science Foundation (BSF) and Grant No. 2109066 from the United States National Science Foundation (NSF), and by the Ministry of Science \& Technology, Israel.

Y.J-T acknowledges financial support from the European Union’s Horizon 2020 research and innovation programme under the Marie Skłodowska-Curie grant agreement No 898633, the MSCA IF Extensions Program of the Spanish National Research Council (CSIC), and the State Agency for Research of the Spanish MCIU through the Center of Excellence Severo Ochoa award to the Instituto de Astrofísica de Andalucía (SEV-2017-0709)

S.B. acknowledges support from the Spanish Ministerio de Ciencia e Innovación through project PGC2018-097585-B-C22 and the Generalitat Valenciana project PROMETEO/2020/085.

I.M. acknowledges financial support from the IAA Severo Ochoa Excellence grant CEX2021-001131-S funded by MCIN/AEI/ 10.13039/501100011033.

L.M. acknowledges support from the grants PRIN-MIUR 2017 WSCC32 and ASI n.2018-23-HH.0.

J.A.F.O. acknowledges the financial support from the Spanish Ministry of Science and Innovation and the European Union -- NextGenerationEU through the Recovery and Resilience Facility project ICTS-MRR-2021-03-CEFCA.

\end{acknowledgements}

\bibliographystyle{aa}

\end{document}